\documentclass[aps,pra,amssymb, amsmath,nobibnotes,nobibnotes,reprint,superscriptaddress,showpacs,showkeys]{revtex4-1}
\usepackage[colorlinks, linkcolor=blue, anchorcolor=blue, citecolor=blue]{hyperref}
\usepackage{bm, amsfonts, mathrsfs}
\usepackage{verbatim, psfrag, times}
\usepackage{graphics, graphicx, color, epsfig}
\usepackage{float}
\usepackage{flafter,natbib}
\usepackage{indentfirst}

\newcommand{\ket}[1]{| #1 \rangle}

\begin{document}

\title{Effect of retardation on the dynamics of entanglement between atoms}

\author{Qurrat-ul-Ain \surname{Gulfam}}
\email{qurrat-ul-ain@mpi-hd.mpg.de}
\affiliation{Max-Planck-Institut f\"ur Kernphysik, Saupfercheckweg 1, 69117 Heidelberg, Deutschland}

\author{Zbigniew \surname{Ficek}}
\email{zficek@kacst.edu.sa}
\affiliation{The National Centre for Mathematics and Physics, KACST, P.O. Box 6086, Riyadh 11442, Saudi Arabia}

\author{J\"org \surname{Evers}}
\email{joerg.evers@mpi-hd.mpg.de}
\affiliation{Max-Planck-Institut f\"ur Kernphysik, Saupfercheckweg 1, 69117 Heidelberg, Deutschland}

\date{\today}

\begin{abstract}
The role of retardation in the entanglement dynamics of two distant atoms interacting with a multi-mode field of a ring cavity is discussed. The retardation is associated with a finite time required for light to travel between the atoms located at a finite distance and between the atoms and the cavity boundaries. We explore features in the concurrence indicative of retardation and show how these features evolve depending on the initial state of the system, distance between the atoms and the number of modes to which the atoms are coupled. In particular, we consider the short-time and the long time dynamics for both the multi- and sub-wavelength distances between the atoms. It is found that the retardation effects can qualitatively modify the entanglement dynamics of the atoms not only at multi- but also at sub-wavelength distances. We follow the temporal evolution of the concurrence and find that at short times of the evolution the retardation induces periodic sudden changes of entanglement. To analyze where the entanglement lies in the space spanned by the state vectors of the system, we introduce the collective Dicke states of the atomic system that explicitly account for the sudden changes as a periodic excitation of the atomic system to the maximally entangled symmetric state. At long times, the retardation gives rise to periodic beats in the concurrence that resemble the phenomenon of collapses and revivals in the Jaynes-Cummings model. In addition, we identify  parameter values and initial conditions at which the atoms remain separable or are entangled without retardation during the entire evolution time, but exhibit the phenomena of sudden birth and sudden death of entanglement when the retardation is included. 
\end{abstract}


\maketitle

\section{Introduction}

Entanglement is one of the most familiar phenomena resulting from the presence of non classical correlations between quantum systems~\cite{Nielsen,ficek}. A large number of studies have demonstrated that entanglement can be created in variety of systems ranging from simple systems such as single photons or atoms to more complex systems such as spin chains or biological samples. The presence of an entanglement between systems has been tested experimentally in various optical experiments. For example, slowly moving atoms can be entangled while passing through a cavity~\cite{PhysRevA.68.033817,*0253-6102-49-4-43}, and the entanglement between the atoms can be detected by probing the atomic state of the atoms after leaving the cavity~\cite{1997,Haroche}. Another common setup is the entanglement of photons obtained from a down-conversion process~\cite{PhysRevA.50.23,*PhysRevA.66.013801,*citeulike:2198147,*Yang:07,*Ueno:12}. In this case, the entanglement between the photons can be verified, e.g., by detecting correlations between their polarizations~\cite{PhysRevLett.91.227901,RevModPhys.73.565}. 

Apart from the issue of creating entanglement, also a detailed analysis of the dynamics of an entangled system is of importance. One motivation for this is the possibility for transferring entanglement between distant quantum systems~\cite{Olmschenk23012009}. Such transfers have become especially interesting since a number of experiments have succeeded in the creation of quantum gates necessary for the implementation of quantum networks~\cite{1464-4266-4-5-309}. 
However, if one examines the dynamics of an entangled system coupled to a network of quantum systems, it becomes apparent that the unavoidable coupling of the systems to the external environment can lead to the irreversible loss of the transferred entanglement. In this connection, one would expect that the coupling of the systems to local environments, with a Markovian assumption of the process, could lead to an exponential decay of the entanglement from its initial value. However, there are some entangled states, particularly those involving at least two excitations, that may decay in an essentially non-exponential manner resulting in the disappearance of the entanglement in a finite time. This effect, known as sudden death of entanglement (SDE), has been studied in a numerous number of papers~\cite{Yu30012009,Ann,ficek,PhysRevA.65.012101,PhysRevA.69.052105,PhysRevLett.93.140404}, and has recently been observed in experiments involving photons~\cite{M.P.Almeida04272007,PhysRevA.78.022322} and atoms~\cite{PhysRevLett.99.180504}. Furthermore, theoretical treatment of the coupling of sub-systems to a common (non-local) environment has predicted that the already destroyed entanglement could suddenly revive~\cite{PhysRevA.74.024304,PhysRevLett.101.080503,Ficek1b,PhysRevA.77.054301,PhysRevA.79.042302} or initially separable systems could become entangled after a finite time, the phenomenon known as sudden birth of entanglement~(SBE).

From experimental point of view, in particular the study of the generation of entanglement in systems confined within optical or microwave cavities~\cite{Berman1994,Haroche,1997,walther2,Rempe2} is of importance. Cavities provide a well-defined mode spectrum and a relatively loss-free environment such that the atom-field interactions can have anomalously large coupling strengths, leading to reversible, non-Markovian type dynamics of the system. As a consequence, the already dead entanglement can revive even if in the equivalent free-space situation no revival is predicted~\cite{Ann,Ikram,Bellomo1,Bellomo2,Bellomo3,0953-4075-40-9-S02,Yu30012009,PhysRevA.81.052107,PhysRevLett.104.250401,PhysRevA.79.042302}. 
However,  calculations based on deriving the master equation for the reduced density operator of two atoms, both placed inside a cavity, frequently assume a large distance between the atoms such that there is no direct interaction between them. At the same moment the treatments assume that each atom influences the other instantaneously. For this, there is no time delay or equivalently no phase difference between the oscillating atomic dipole moments resulting in an effective coupling between the atoms independent of their distance. 
It turns out that these non-retarded models to physical systems are suitable if the atoms interact with a single cavity mode. 
A more interesting parameter regime arises if the atom couples to a large number of cavity modes. In this case, retardation effects become important~\cite{Milonni,PhysRevA.2.1730,Giessen,Meystre}. These effects are associated with a finite time required for light to travel, e.g., from the atom to the boundary of the cavity and back to the atom after being reflected from the cavity mirror. This leads to the interference between instantaneously emitted photon and the retarded waves that are reflected from the cavity walls. 

In an early study Milonni and Knight~\cite{Milonni,PhysRevA.2.1730} discussed the effect of the retardation on the collective behavior of two atoms. They demonstrated that that retardation effects in the interaction between two atoms in free space become important for distances larger than the half-wavelength of the field. Recent studies of the interaction of atoms with multi-mode cavities have predicted strong non-Markovian and retardation effects in the population dynamics~\cite{Giessen,Meystre}. The multi-mode cavity field can be treated as a small environment to the atoms~\cite{shu}. This leads to a spatial modulation of the field amplitude which significantly alters the nature of the interaction between the atoms and the field.

While most of the studies on SDE and SBE assumed the Markovian approximation such that a backaction of the environment on the atoms is effectively excluded, recently, however, also the non-Markovian case has received considerable attention~\cite{Ann,Ikram,Bellomo1,Bellomo2,Bellomo3,0953-4075-40-9-S02,Yu30012009,PhysRevA.81.052107,PhysRevLett.104.250401,PhysRevA.79.042302}. In particular those explicitly taking into account the distance between the particles are of relevance~\cite{q-inf,q-inf2,q-inf3}. In these  works it was shown that for certain initial states, the distance between the qubits can qualitatively change the entanglement dynamics. For example, depending on the distance, SDE and SBE can occur or not. However, these works made use of a continuum of environmental modes by integrating over all wave vectors $\vec{k}$. This raises the question about the entanglement dynamics of atoms in multi-mode cavities with a set of discrete field modes, which are known to exhibit strong non-Markovian and retardation effects in the population dynamics~\cite{Meystre}. 

In this paper we investigate the effect of retardation on the generation and dynamics of entanglement between two two-level atoms located inside a ring cavity. The model studied requires that we develop a multi-mode theory of the interaction of the atoms with the cavity field. The goal then is to trace the time evolution of the concurrence in the case of single or double excitations present in the system. We show that the quantum nature of the cavity field crucially affects the generation of entanglement in the system. In the course of the calculation we observe that the retardation effects do play a significant role in the creation of entanglement between the atoms. Certain transient effects such as abrupt kinks in the time evolution of the populations and the concurrence occur. The kinks reflect the effects of multiple photon exchange between the atoms and appear at intervals corresponding to the multiplets of time required for the photon to travel between the atoms or to take the round trip in the cavity. The effect of the retardation on the phenomena of sudden death, revival and sudden birth of entanglement is also discussed. 
In particular, we identify parameters and initial conditions, in which the atoms remain separable without retardation throughout the entire evolution time, but exhibit sudden birth and death of entanglement with retardation, and vice versa. Both, the short-time and the long time dynamics are analyzed, and we also study time-averaged concurrences. We also study  the distance dependence on two scales: First, in integer multiples of the wavelength, corresponding to different positions in a periodic potential, and second on a sub-wavelength scale.

We begin in Sec.~\ref{theo} by introducing the model and derive the equations of motion for the probability amplitudes in two cases of single and double excitations present in the system. These equations are obtained by considering a multi-mode rather than a single-mode interaction of the atoms with the cavity field. Then, in Sec.~\ref{concu} we apply the solutions for the probability amplitudes to the problem of the time evolution of the populations and the concurrence. Throughout, we assume that the atoms interact with a finite number of the cavity modes. The numerical results for various special cases of the time evolution of the concurrence are illustrated in Sec.~\ref{results}. We also present there the qualitative discussion of the short and long time behaviors of the concurrence. Finally, in Sec.~\ref{summary} we summarize our results.

\section{The model}\label{theo}

\begin{figure}[b]
\includegraphics[width=0.8\linewidth ]{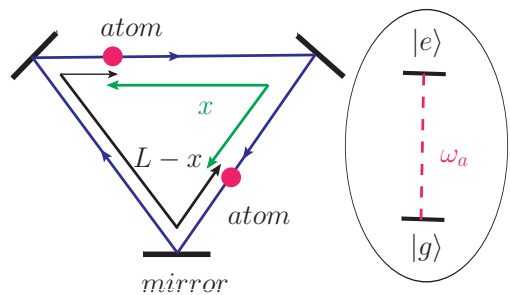}
\caption{\label{system}(Color online) Schematic diagram of the system considered. Two identical two-level atoms are at fixed positions, distant $x$ from each other, located inside a one-dimensional ring cavity of the round trip path $L$. The internal structure of each atom is shown in the right inset. The atoms are modeled as two-level systems with the excited $|e_{i}\rangle$ and ground $|g_{i}\rangle$ states separated by the transition frequency $\omega_a$.}
\end{figure}

We consider two identical atoms, located inside a ring cavity at fixed positions $\vec{x}_1$ and $\vec{x}_2$, with distance $|\vec{x}_2-\vec{x}_1|=x$. The atoms are modeled as two-level systems with excited state~$|e_{i}\rangle$ and ground state $|g_{i}\rangle \, (i\in\{1,2\})$ separated by energy~$\hbar\omega_{a}$, as shown in Fig.~\ref{system}. 
The cavity is considered as a multi-mode cavity with frequency difference between adjacent modes (free spectral range) such that multiple modes are supported within the atomic resonance line width.  The consideration of several rather than a single mode in the interaction of the atoms with the cavity field will be found crucial for the occurrence of retardation in the radiative coupling between the atoms.

\subsection{Hamiltonian of the system}

The Hamiltonian of the atoms interacting with the common electromagnetic field of the ring cavity can be written as~\cite{Meystre}
\begin{equation}
H = H_a+H_f+H_{af} ,\label{hamiltonian}
\end{equation}
where 
\begin{equation}
H_a =\sum_{j=1}^2 \hbar \omega_a S^{+}_{j} S^{-}_{j} \label{eq2}
\end{equation}
is the free Hamiltonian of the atoms,
\begin{equation}
H_f=\hbar\sum_{\mu} \omega_n a_{\mu}^{\dagger} a_{\mu} \label{eq3}
\end{equation}
is the free Hamiltonian of the cavity field, and 
\begin{equation}
H_{af} = -\vec{D}_1\cdot\vec{E}(\vec{x}_1)-\vec{D}_2\cdot\vec{E}(\vec{x}_2) \label{eq4}
\end{equation}
is the interaction between the atoms and the cavity field, written in the electric dipole approximation.

In the Hamiltonian (\ref{hamiltonian}), the atoms are represented by the transition dipole moment operators
\begin{equation}
\vec{D}_{j}= \vec{d}_{j}S^{+}_{j} + \vec{d}_{j}^{\ast}S^{-}_{j} ,\label{eq5}
\end{equation}
where $S^{+}_{j}=|e_j\rangle \langle g_j|$ and $S^{-}_{j}=|g_j\rangle \langle e_j|$ are respectively the dipole raising and lowering operators of the atom $j$, and $\vec{d}_{j}=\langle g_{j}|\vec{D}_{j}|e_{j}\rangle$ is the dipole matrix element of the atomic transition.

The cavity field is represented by the creation $a_{\mu}^{\dagger}$ and annihilation $a_{\mu}$ operators in which the subscript $\mu$ indicates the particular set of the cavity plane-wave modes $\mu=\{\vec{k}_n,l\}$ of the wave number $k_n =\omega_n/c$, frequency $\omega_{n}$ and polarization~$l$, to which the atoms are coupled. 

The cavity field at position $\vec{x}$ can be given in the plane-wave mode expansion as
\begin{equation}
\vec{E}(\vec{x}) = i \sum_{\mu}\mathcal{E}_{\mu}\left(a_{\mu}{\rm e}^{i\vec{k}_n\cdot\vec{x}}\hat{e}_l - \rm{H.c.}\right) ,\label{eq6}
\end{equation}
where 
\begin{equation}
\mathcal{E}_{\mu} =\sqrt{\dfrac{\hbar\omega_n}{2\epsilon_0 L}},\label{eq7}
\end{equation}
is the electric field amplitude of the $n$th mode, $\omega_n=2 \pi n c/L$ is in the frequency of the modes set by the periodic boundary conditions of the ring cavity, and $\hat{e}_{l}$ is the unit polarization vector of the mode $\mu$. 

After substituting Eqs.~(\ref{eq5}) and (\ref{eq6}) into Eq.~(\ref{eq4}), and retaining only the terms which play a dominant role in the rotating wave approximation, the interaction Hamiltonian takes the form 
\begin{equation}
H_{af} = i\hbar \sum_{j=1}^{2}\sum_{\mu}\left[g_{\mu}(\vec{x}_j)a_{\mu}S^{+}_{j} - \rm{H.c.}\right] ,\label{eq8}
\end{equation}
where 
\begin{equation}
g_{\mu}(\vec{x}_j) = \frac{\mathcal{E}_{\mu}}{\hbar}\left(\vec{d}_{j}\cdot \hat{e}_{l}\right) {\rm e}^{i \vec{k}_n\cdot\vec{x}_j} \label{eq9}
\end{equation}
is the position-dependent Rabi frequency which determines the strength of the coupling of the $j$th atom with the mode $\mu$ of the cavity field.

Our objective is to find effects of the retardation in the interaction of the atoms with the multi-mode cavity field on the evolution of the system. We are in particular interested in the effect of the retardation on the creation of entanglement between the atoms. Two cases will be studied, with the system initially (1) in a single excitation state, and (2) in a double excitation state. Before going into detailed calculations, we first briefly explain how retardation effects are incorporated in our calculations.

\subsection{Origin of the retardation}

The atom-cavity system exhibits retardation effects if its dynamics is affected by the finite propagation time of the light. In our model, two effects need to be distinguished~\cite{Meystre}. First, an atom embedded in a larger cavity initially evolves as in free space, but after a finite time of order $L/c$ (and integer multiples thereof) reacts to the presence of the cavity with a sudden kink in the time evolution. Speaking pictorially, this time is required for a photon emitted by the atom to cycle through the cavity and be reabsorbed by the same atom. This gives rise to retardation effects which occur already for a single atom in the cavity. The second effect is due to the interaction of two atoms in the cavity. Here, the retardation occurs because of the finite time required for a photon to travel between the two atoms. 

\begin{figure}[h]
\includegraphics[width=0.5\linewidth ]{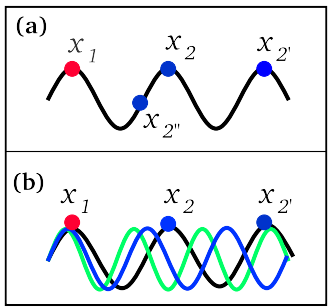}
\caption{(Color online) Atoms represented by colored blobs inside the cavity field. $x_1$ denotes the position of atom 1 which is placed at the first anti node. (a) Single mode of the cavity field with three different positions for atom 2. (b) Two additional modes shown in blue and green are taken into account.}
\label{modes}
\end{figure}

In the following, we provide an intuitive picture how the distance information required for the two different retardation effects enters our model.  This discussion will be made more precise in Sec.~\ref{case-single}, where we identify the origin of the two types of retardation in the equations of motion governing the atom-cavity system.

A typical representation of the interaction of an atom with a cavity field is illustrated in Fig.~\ref{modes}. 
Evaluating the electric field operator (\ref{eq6}) in the Heisenberg picture to include the time evolution, we find
\begin{equation}
\vec{E}(\vec{x}) = i \sum_{\mu}\mathcal{E}_{\mu}\left(a_{\mu}{\rm e}^{i\varphi_n(\vec{x},t)}\hat{e}_l - \rm{H.c.}\right)\,.
\end{equation}
The time and space dependence enters via the phases  $\varphi_n(\vec x,t) = \vec{k}_n \vec{x} - \omega_n t = \omega_n (x/c - t)$ of the different modes. In the last step  and throughout this section, we assume a one-dimensional problem, and thus $\vec{k}_n \vec{x} = k_n x$ for simplicity.

Let us first discuss a simplified model, the interaction of the atom with a single mode only, say the central cavity mode $\mu =0$, as shown in Fig.~\ref{modes}(a). Suppose now that atom $1$ represented by a red blob is located at position $x_{1}$, whereas atom $2$, represented by a blue blob can be located at three different positions $x_{2}, x_{2^{\prime}}$ and~$x_{2^{\prime\prime}}$. It is easily verified that if $x$ is displaced by $\delta x$ such that $k_{0}\cdot\delta x = 2\pi n$, where $n$ is an integer, then the phase of the single mode remains unchanged,   $\varphi_0(x,t) = \varphi_0(x+\delta x, t)$. This implies that the electric field operator has the same value at positions differing by an integer multiple of the cavity wavelength. Hence, the interaction Hamiltonian $H_{af}$ remains the same if the distance between the two atoms is changed by an integer multiple of the wavelength. In this sense, the system dynamics is independent of the distance, and thus the system itself cannot exhibit effects of retardation if only a single mode is considered.

It should be noted that the interaction Hamiltonian  still depends on the relative distance between the atoms. For atomic separations $|x_2-x_1|$ and $|x_{2^{\prime}}-x_1|$ that are equal to multiple integer of the wavelength, $g_{0}(x_{1})=g_{0}(x_{2})=g_{0}(x_{2^{\prime}})$, since at these separations the Rabi frequencies have the same value. In other words, the atoms are coupled equally to the field mode. But for separations between the atoms that do not satisfy the periodicity condition, such as $|x_{2^{\prime\prime}}-x_1|$, the atoms experience different amplitudes and phases of the field. As a consequence, the interaction $H_{af}$ is modified due to to a change of the Rabi frequency.  However, this variation of the interaction Hamiltonian with the inter atomic distance in the single mode field has nothing to do with retardation. 

Let us now assume that apart from the central mode, there are additional cavity modes taking part in the interaction with the atoms, as illustrated in Fig.~\ref{modes}(b). Suppose that at point~$x_1$, all modes have the same phase ($\varphi_n(x_1,t) = \varphi_0$ for all $n$), which we for simplicity assume to be zero. Then, the position $x$ entering the field operator can be interpreted as the distance $\delta x = x_2 - x_1$ between the two atoms. Due to their different wave numbers~$k_{n}$, at point~$x_2$, the modes typically have different phases, i.e., $\varphi_n(x_2,t) \neq \varphi_m(x_2,t)$ for $n\neq m$.  In other words, the modes are shifted or ``retarded" with respect to each other. This means that the atom at position $x_2$ experiences the field emitted by atom 1 into the different cavity modes with different relative phases, such that the response of atom 2 in the limit of large mode number averages out. However, at specific times, all modes can evolve in phase again. From the definition of $\varphi_n(x,t)$, it is clear that this happens at times $x/c$, i.e., exactly the times corresponding to the flight time of light between the two atoms. At this instance in time, the cavity modes act in phase onto atom 2, such that a sudden response is observed. This is the origin for the retardation effects of the second type. 

In contrast, the retardation effects of the first type are embedded in the quantization of the cavity modes frequency spectrum. The frequency spacings are such that the phases at times separated by integer multiples of $L/c$ are different by an integer multiple of $2\pi$, as $\omega_n L/c = 2\pi n$. While at most times the different modes are out of phase because of the different free evolution frequencies, at times equal to integer multiples of $L/c$ all modes are in phase again, and a sudden response of the atom appears. From this interpretation, it is apparent that this first type of retardation already occurs for a single atom.

A straightforward combination of both arguments also explains the retardation of the second type at times $(L-x)/c$. Furthermore, in the subsequent evolution of the atoms also combinations of the different retardation time intervals can occur, as we will see in the numerical analysis.

Thus, retardation effects are expected to play an important role in the interaction of atoms as soon as multiple modes with different wave numbers interact with the atoms. This can also be related to the Fourier relation of position and momentum space. The small frequency distribution of a single mode gives rise to a large distribution or uncertainty in the position. In contrast, a broad frequency distribution or many modes allow to precisely determine the position.

\section{Excitation probabilities and concurrence}

We are interested in determining the effect of retardation on the time evolution of the system initially prepared in separable states of single and double excitations. In particular, we shall discuss the subsequent time-dependent behavior of the excitation probabilities and the concurrence. 

The time evolution of the system is governed by the Schr\"odinger equation, which in the interaction picture is given~by 
\begin{equation}
i\hbar\frac{\partial |\psi(t)\rangle}{\partial t} = H_{af}|\psi(t)\rangle .\label{eq11}
\end{equation}
We will consider two particular classes of initial conditions. In the first, we assume that at $t=0$ the system was in a single-excitation state. In the second class, we assume that initially two excitations were present in the system. In both cases, the (single or double) excitation is initially either present in the atoms or in the cavity, or in a superposition of atoms and cavity.

\subsection{\label{case-single}The case of single excitation}

If we take for the initial state of the system a single-excitation state, then the time-dependent state vector of two atoms coupled to a multi-mode field can be written as
\begin{align}
|\psi(t)\rangle &= b_1(t)|e_{1} g_{2} \{0\}_{\mu}\rangle + b_2(t)|g_{1} e_{2} \{0\}_{\mu}\rangle \nonumber\\
&+\sum_{\mu}b_{\mu}(t)|g_{1}g_{2}\{1\}_{\mu}\rangle ,\label{eq13}
\end{align}
where $\{1\}_{\mu}$ denotes the state of the cavity modes with  a single excitation present in the mode $\mu$ and zero occupation numbers for all the remaining modes.

The time evolution of the state vector is determined by the Schr\"odinger equation~(\ref{eq11}) which transforms it into three coupled equations of motion for the probability amplitudes
\begin{subequations}
\begin{align}
\label{eqofmotion1}
\dot{b}_{j}(t)=&\sum_{\mu}g_{\mu j} b_{\mu}(t) ,\quad j \in \{1,2\} ,\\ \label{eqofmotion3}
\dot{b}_{\mu}(t)=&-i\Delta_{\mu} b_{\mu}(t) - \sum_{j=1}^{2}g^*_{\mu j}b_{j}(t) ,
\end{align}
\end{subequations}
where $\Delta_{\mu}=\omega_{\mu}-\omega_a$ is the detuning of the cavity mode frequency $\omega_{\mu}$ from the atomic transition frequency $\omega_a$ coinciding with the central cavity mode frequency, and we have simplified the notation $g_{\mu i}\equiv g_{\mu}(\vec{x}_i)$.

The formal integration of Eq.~(\ref{eqofmotion3}) gives
\begin{align}
b_{\mu}(t) = b_{\mu}(0){\rm e}^{-i\Delta_{\mu}t} - \sum_{j=1}^{2}g^{\ast}_{\mu j}\int_{0}^{t}dt^{\prime}b_{j}(t^{\prime}){\rm e}^{-i\Delta_{\mu}(t-t^{\prime})}  ,\label{eq16}
\end{align}
and when this relation is substituted into Eq.~(\ref{eqofmotion1}), we find
\begin{align}
\dot{b}_{j}(t) &= \sum_{\mu}g_{\mu j} b_{\mu}(0){\rm e}^{-i\Delta_{\mu}t} \nonumber\\
&- \sum_{j^{\prime}=1}^{2}\int_{0}^{t}dt^{\prime}\sum_{\mu}g_{\mu j}g^{\ast}_{\mu j^{\prime}}b_{j^{\prime}}(t-t^{\prime}){\rm e}^{-i\Delta_{\mu}t^{\prime}} .\label{eq17}
\end{align}

At this point, the two types of retardation effects are fully visible also from the analytical expressions. The first type leading to retardation effects at times equal to integer multiples of the cavity round trip time $L/c$ can be understood in terms of the detunings $\Delta_\mu$. Since the detunings $\Delta_{\mu}$ differ by integer multiples of $2 \pi c/L$, the phases $\Delta_\mu t$ will be multiples of $2\pi$ for all modes $\mu$ simultaneously at times $t$ equal to integer multiples of $L/c$. Then, the system response will exhibit sharp peaks due to the constructive interference of all modes. At other times, the different phase factors of the various modes do not add up constructively. This discrete response is independent of the coupling between the atoms, and could be observed even if only a single atom is present inside the cavity. 

The second type of retardation is due to the interaction between the atoms, i.e., the second part of Eq.~(\ref{eq17}).  The coupling constants $g_{\mu j}g_{\mu j'}^*$ together with the detuning phase lead to phase contributions $i [k_\mu (x_1-x_2) - \omega_\mu t] = i \omega_\mu [ (x_j-x_j')/c - t]$ for the different modes. Again, constructive interference is obtained, but in this case at a time corresponding to the flight time $x/c$ between the two atoms. A similar argument also explains constructive interference at time $(L-x)/c$.

Subsequent iterations of the two types of retardation lead to kinks in the system evolution also at times arising from combinations of the two effects. Obviously, constructive interference can only lead to sharp change in the system response if many different frequency components contribute, i.e., if the system couples to many cavity modes. In the extreme case of free space, $L\to \infty$, and the first type of retardation cannot occur at a finite time. But the retardation in the coupling between the two atoms is still present in the free-space limit, and must be considered, e.g., in calculating the dipole-dipole coupling between atoms in free space.

\subsection{The case of double excitation}

If initially the system was in a double excitation state, then the state vector can be written as
\begin{align}
|\tilde{\psi}(t)\rangle &= b_{12}(t)|e_{1}e_{2}\{0\}_{\mu}\rangle +\sum_{\alpha}b_{\alpha1}(t)|e_{1}g_{2}\{1\}_{\alpha}\rangle \nonumber\\
&+\sum_{\alpha}b_{\alpha2}(t)|g_{1}e_{2}\{1\}_{\alpha}\rangle +\sum_{\alpha}b_{\alpha\alpha}(t)|g_{1}g_{2}\{2\}_{\alpha}\rangle\nonumber\\
&+\sum_{\alpha>\beta}b_{\alpha\beta}(t)|g_{1}g_{2}\{1\}_{\alpha}\{1\}_{\beta}\rangle ,\label{extstatevec}
\end{align}
where $\{2\}_\alpha$ denotes the state of the cavity modes with double excitation of the mode $\alpha$ and zero occupation numbers for all the remaining modes. 

The Schr\"odinger equation transforms the state vector (\ref{extstatevec}) into the following set of coupled equations of motion for the probability amplitudes
\begin{align}
\dot{b}_{12}(t) =& \sum_{j\neq j^{\prime}=1}^{2}\sum_{\alpha} g_{\alpha j} b_{\alpha j^{\prime}}(t) ,\nonumber\\
\dot{b}_{\alpha j}(t) =& -i\Delta_{\alpha}b_{\alpha j}(t) -g^{*}_{\alpha j^{\prime}}b_{12}(t)+\sqrt{2}g_{\alpha j}b_{\alpha\alpha}(t) \nonumber\\
+&\sum_{\beta>\alpha}g_{\beta j}b_{\beta\alpha}(t)\!+\!\sum_{\beta<\alpha}g_{\beta j}b_{\alpha\beta}(t) ,\quad (j\neq j^{\prime}\in\{1,2\}) ,\nonumber\\ 
\dot{b}_{\alpha\beta}(t) =& -i(\Delta_{\alpha}\!+\!\Delta_{\beta})b_{\alpha\beta}(t)\!-\!\sum_{j=1}^{2}\!\left[g^{*}_{\alpha j}b_{\beta j}(t)\!+\!g^{\ast}_{\beta j}b_{\alpha j}(t)\right] ,\nonumber\\
\dot{b}_{\alpha\alpha}(t) =& -2i\Delta_{\alpha}b_{\alpha\alpha}(t)-\sqrt{2}\sum_{j=1}^{2}g^{*}_{\alpha j}b_{\alpha j}(t) .\label{eqofmotion2}
\end{align}
The case of double excitation is described by a complicated set of equations of motion. It involves probability amplitudes of the states with the excitation redistributed over two cavity modes, $b_{\alpha\beta}(t)$, as well as states with the excitation occupying the same mode $b_{\alpha\alpha}(t)$.

\subsection{Concurrence}\label{concu}

We are mainly interested in studying the retardation effects on the entanglement dynamics between the two atoms that are coupled to the multi mode vacuum field inside the cavity. The dynamics of the atoms are determined by the reduced density matrix $\rho$ that is obtained by tracing the density matrix of the total system over the field degrees of freedom. We then exploit concurrence introduced by Wootters~\cite{PhysRevLett.80.2245,*PhysRevLett.78.5022}, which is a widely accepted measure of entanglement between two qubits, and is defined by
\begin{equation}
\label{maxim}
C =\rm{max}\{0,\sqrt{\lambda_1}-\sqrt{\lambda_2}-\sqrt{\lambda_3}-\sqrt{\lambda_4}\} ,
\end{equation}
where $\lambda_i$ are the eigenvalues (in descending order) of the Hermitian matrix $R=\rho\tilde{\rho}$ in which $\tilde{\rho}$ is given by
\begin{equation}
\tilde{\rho} = \sigma_y\otimes\sigma_y\rho^{\ast}\sigma_y\otimes\sigma_y.
\end{equation}
and $\sigma_y$ is a Pauli matrix. The concurrence ranges between 0 and 1. If the two atoms are maximally entangled, the concurrence evaluates to unity whereas, if they are completely disentangled, $C=0$. 

The usual way is to express the concurrence in the basis of the product states of the two-atom system, i.e., $|1\rangle=|e_{1}e_{2}\rangle,|2\rangle=|e_{1}g_{2}\rangle,|3\rangle=|g_{1}e_{2}\rangle, |4\rangle=|g_{1}g_{2}\rangle$. In this basis, the concurrence takes the form~\cite{Ikram}
\begin{align}
C(t) =& \, 2\, {\rm max}\big\{0,|\rho_{23}(t)|-\sqrt{\rho_{44}(t)\rho_{11}(t)} ,\nonumber\\
&|\rho_{14}(t)| -\sqrt{\rho_{22}(t)\rho_{33}(t)}\big\} .\label{ikram}
\end{align}
There are two terms contributing to the concurrence, one resulting from the presence of the one-photon coherence $|\rho_{23}(t)|$ and the other from the two-photon coherence $|\rho_{14}(t)|$. It is interesting that these two contributions complement each other. In the single excitation case, $\rho_{11}=0, \rho_{14}=0$, and then the expression for the concurrence (denoted in this case by \textbf{C}) reduces to 
\begin{equation}
\textbf{C}(t) = 2\, {\rm{max}}\{0,|\rho_{23}(t)|\} = 2\, {\rm{max}}\{0,|b^*_1(t)\,b_2(t)|\} .\label{Conc.}
\end{equation}
It shows that in the single excitation case it is sufficient for $|\rho_{23}(t)|$ to be different from zero to create entanglement between the atoms. In this sense, entanglement is equivalent to atomic coherence in this case.

The situation is quite different when two excitations are present in the system. But surprisingly, tracing the density matrix of the system over the field degrees of freedom results in an expression for the concurrence that does not involve the two-photon coherence $\rho_{14}(t)$. To see this more explicitly, we calculate the density matrix $\rho_{T}$ associated with the two-excitation state Eq.~(\ref{extstatevec}), and find
\begin{align}
\rho_{T} &=\, |b_{12}(t)|^2|e_{1}e_{2}\{0\}_{\mu}\rangle\langle e_{1}e_{2}\{0\}_{\mu}| \nonumber\\
&+\sum_{\alpha}|b_{\alpha1}(t)|^2|e_{1}g_{2}\{1\}_{\alpha}\rangle\langle e_{1}g_{2}\{1\}_{\alpha}|\nonumber \\
&+\sum_{\alpha}b_{\alpha1}(t)b^*_{\alpha2}(t)|e_{1}g_{2}\{1\}_{\alpha}\rangle\langle g_{1}e_{2}\{1\}_{\alpha}|\nonumber \\
&+\sum_{\alpha}b_{\alpha2}(t)b^*_{\alpha1}(t)|g_{1}e_{2}\{1\}_{\alpha}\rangle\langle  e_{1}g_{2}\{1\}_{\alpha}| \nonumber \\
&+\sum_{\alpha}|b_{\alpha2}(t)|^2|g_{1}e_{2}\{1\}_{\alpha}\rangle\langle g_{1}e_{2}\{1\}_{\alpha}| \nonumber \\
&+\sum_{\alpha>\beta}|b_{\alpha\beta}(t)|^2|g_{1}g_{2}\{1\}_{\alpha}\{1\}_{\beta}\rangle\langle g_{1}g_{2}\{1\}_{\alpha}\{1\}_{\beta} | \nonumber \\
&+\sum_{\alpha}|b_{\alpha\alpha}(t)|^2|g_{1}g_{2}\{2\}_{\alpha}\rangle\langle g_{1}g_{2}\{2\}_{\alpha}| + ND ,\label{eq23}
\end{align}
where $ND$ stands for the sum of all off-diagonal terms in the field modes which vanish in tracing over the cavity modes. Then, by taking trace of the density matrix $\rho_{T}$ over the cavity modes, we arrive at the following reduced density matrix of the two atoms
\begin{align}
\rho &=\, |b(t)|^2|1\rangle\langle 1| +\sum_{\alpha} |b_{\alpha1}(t)|^2  \: |2\rangle\langle 2| +\sum_{\alpha}|b_{\alpha2}(t)|^2\:|3\rangle\langle 3| \nonumber \\
&+\sum_{\alpha}b_{\alpha1}(t)b^*_{\alpha2}(t)\:|2\rangle\langle 3 |
+\sum_{\alpha}b_{\alpha2}(t)b^*_{\alpha1}(t)\:|3 \rangle\langle  2 | \nonumber \\
&+  \left(\sum_{\alpha>\beta}|b_{\alpha,\beta}(t)|^2 +\sum_{\alpha}|b_{\alpha,\alpha}(t)|^2 \right)\:|4\rangle\langle 4 | .
\end{align}
It is clear that tracing out the field modes results in the density matrix with $\rho_{14}=0$. In this case, the concurrence denoted by $\mathcal{C}$ takes the form
\begin{align}
\mathcal{C}(t) = 2\, {\rm max}\big\{0,|\rho_{23}(t)|-\sqrt{\rho_{44}(t)\rho_{11}(t)}\big\} ,\label{eq25}
\end{align}
which in terms of the probability amplitudes can be written as
\begin{align}
\mathcal{C}(t) = {\rm{max}}\{0,\mathcal{C}_1(t)\} ,\label{eq26}
\end{align}
where
\begin{align}
\mathcal{C}_1(t) &= 2\sum_{\alpha}|b^*_{\alpha2}(t)b_{\alpha1}(t)| \nonumber\\
&-|b(t)|\sqrt{\sum_{\alpha>\beta}|b_{\alpha,\beta}(t)|^2+\sum_{\alpha}|b_{\alpha,\alpha}(t)|^2} . \label{eq27}
\end{align}
Similar to the single excitation case, the concurrence depends on the coherence $\rho_{23}(t)$. However, in the presence of two excitations in the system, the condition for a nonzero concurrence of $|\rho_{23}(t)|\neq 0$ is a necessary one, it is not in general sufficient one, since there is a subtle condition of the coherence to be larger than a threshold value of $\sqrt{\rho_{44}(t)\rho_{11}(t)}$. Thus, the presence of the two excitations in the system introduces a threshold for the coherence above which the entanglement between the atoms could occur. Needless to say, the first term in Eq.~(\ref{eq27}), $|b^*_{\alpha2}(t)b_{\alpha1}(t)|$, must be different from zero and exceed the second term numerically for the concurrence to be positive.  

We should point out here that the involvement of only the one-photon coherence in the concurrence of the double excitation case is a direct consequence of the quantum nature of the field. The definite total excitation number entangles the excitation number of the atoms uniquely to the excitation number of the cavity. If the cavity is projected into particular excitation number channels with classical probabilities not allowing for quantum superpositions in tracing over the cavity modes, due to this entanglement, also the atoms are projected into the corresponding excitation number subspaces. This rules out coherence or even entanglement between atomic states of different excitation number.  This situation was treated by Yonac {\it et al.}~\cite{yye06,yye07}, who showed that in the case of a two mode cavity, $\alpha\in\{1,2\}$, no coherence and equivalently no entanglement can be found in a system determined by the double excitation state (\ref{extstatevec}). 

The coherence could be present if one includes an auxiliary state $|g_{1}g_{2}\{0\}_{\alpha}\{0\}_{\beta}\rangle$, the ground state with no excitation, to the state (\ref{extstatevec}). Then, the total excitation number would not be fixed, and there would be no definite entanglement between the atom and cavity excitation numbers.
Alternatively, if the photon number states in Eq.~(\ref{eq23}) were replaced by a classical field amplitude, for example, by a coherent state $\ket{\alpha}$, one could then arrive at the concurrence involving the two-photon coherence $\rho_{14}$. Thus, the condition for entanglement based on $\rho_{14}$ would become relevant. It is easy to see, replacing in Eq.~(\ref{eq23}) the photon number states $\ket{\{n\}_{\mu}}$ by the coherent state $\ket\alpha$, we obtain a state vector
\begin{align}
\label{extstatevec-coherent}
|\hat{\psi}(t)\rangle &= b_{12}(t)|e_{1}e_{2}\alpha\rangle +\sum_{\alpha}b_{\alpha1}(t)|e_{1}g_{2}\alpha\rangle \nonumber\\
&+\sum_{\alpha}b_{\alpha2}(t)|g_{1}e_{2}\alpha\rangle +\sum_{\alpha}b_{\alpha\alpha}(t)|g_{1}g_{2}\alpha\rangle.
\end{align}
Using $\hat{\psi}(t)$ from Eq.~(\ref{extstatevec-coherent}), one can calculate the density matrix $\rho^{cl}$. Now
\begin{align}
\rho^{cl}_{14}=&\langle 1|\rho^{cl}| 4 \rangle\nonumber\\
=& b_{12}^*(t)b_{\alpha\alpha}(t)\langle e_{1}e_{2}\alpha|\hat{\psi}(t)\rangle\langle \hat{\psi}(t)|g_{1}g_{2}\alpha\rangle\nonumber\\
=& |b_{12}^*(t)|^2 |b_{\alpha\alpha}(t)|^2.
\end{align}
It is seen that the resulting density matrix element containing contributions from the two-photon coherences no more vanishes.  This is consistent with our interpretation, as a coherent state has a distribution of photon numbers rather than a well-defined occupation as a Fock state.

\section{Results and discussion}\label{results}

Having discussed the general features of the concurrence, we now turn to analyze the transient behavior of the populations and concurrence for initial conditions in which the atoms are prepared in separable single or double excitation states and for an initial condition in which the atoms are initially in a partially or a maximally entangled state. We shall allow for an arbitrary atomic spacing and length of the cavity, but we limit the discussion to situations in which the central cavity mode is on resonance with the transition frequency of the atoms, i.e., $\omega_{0}=\omega_a$. Also, since the coupling of the atoms to the cavity modes decreases with increasing detuning, we take into account in the numerical calculation a finite number of cavity modes distributed about $\omega_{0}$ with a frequency range on the order of several atomic line widths. The required number of modes depends on the cavity length $L$, as the distance between the adjacent cavity modes decreases with increasing~$L$. Thus, the number of the cavity modes to which the atoms can be coupled increases with an increasing~$L$. 

Equations (\ref{Conc.}) and (\ref{eq25}) for the concurrence are functions of several parameters: the atomic spacing $x$, the detuning of the cavity modes from the atomic transition frequency $\Delta_{\mu}$, the number of the cavity modes $N$ to which the atoms are coupled, the coupling strength of the atoms to the cavity modes $g_{\mu i}$, the cavity length $L$, and the time $t$. For fixed $N$ and $L$, one can obtain time evolution of the concurrence by monitoring the populations of the atomic states and coherence between the atoms as a function of $t$. Alternatively, one can monitor the time evolution of the populations of the collective states of the two-atom system. In the following, we give illustrative figures of both on a short time and a long time behavior of the concurrence.

\subsection{Effects of retardation on the population dynamics}

Before discussing the effects of retardation on the transient properties of the concurrence, it is important to understand the transient behavior of the populations of the single-excitation case. Transient excitation probabilities are first studied for arbitrary initial conditions for the atomic and the collective states of the system. The effects of retardation on the population dynamics were studied by Goldstein and Meystre~\cite{Meystre}. However, these calculations were not specifically oriented towards studying the transient properties of the collective states of the system which, as we shall see below, are very useful for the interpretation of the entanglement dynamics of the atoms.

To calculate the population dynamics, we solve numerically the set of coupled equations for the probability amplitudes, Eqs.~(\ref{eqofmotion1})-(\ref{eqofmotion3}), assuming that the atoms were prepared at time $t=0$ in a product state 
\begin{equation}
|\psi(0)\rangle = |e_{1}\rangle\otimes |g_{2}\rangle \otimes |\{0\}_{\mu}\rangle \equiv |e_{1}g_{2}\{0\}_{\mu}\rangle ,\label{eq12}
\end{equation}
where $|\{0\}_{\mu}\rangle$ denotes the product state vector of the cavity modes with zero occupation numbers for all the modes~$\mu$. The initial condition (\ref{eq12}) corresponds to $b_{1}(0)=1$ and $b_{2}(0)=b_{\mu}(0)=0$. We then compute the time evolution of the excitation probabilities~$|b_1(t)|^2$ and $|b_2(t)|^2$ of the atoms.
\begin{figure}[t]
\includegraphics[width=0.8\linewidth ]{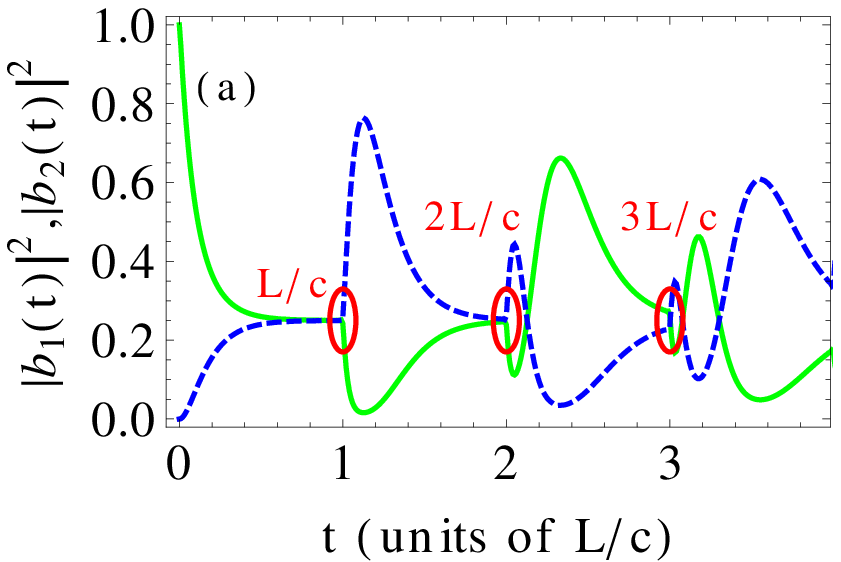}
\includegraphics[width=0.8\linewidth ]{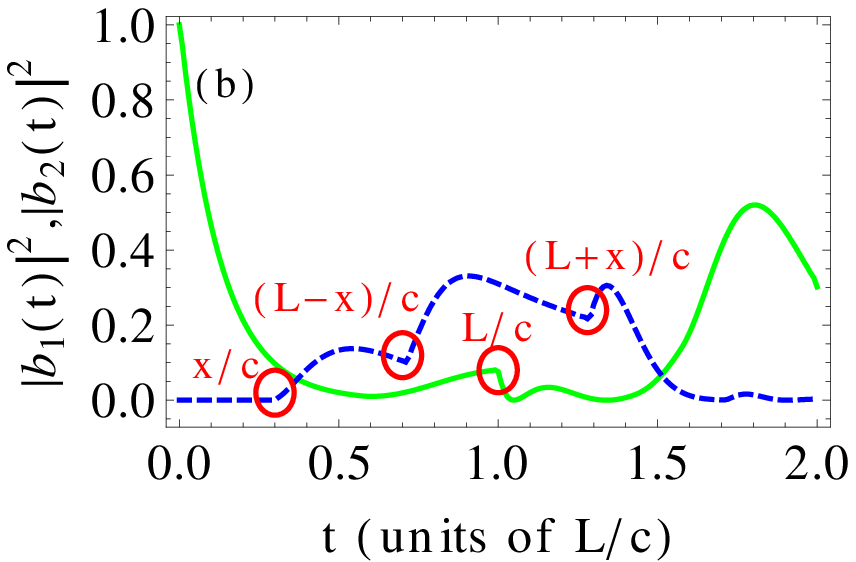}
\caption{\label{population}(Color online) The time variation of the probabilities $|b_1(t)|^2$ (solid green line) and $|b_2(t)|^2$ (dashed blue line) for $L=3.48\times 10^3\lambda_a$, $N=99, \omega_a=1.11\times 10^4\Omega_{0}$, where $\Omega_{0}$ is the vacuum Rabi frequency of the central resonant mode, and atomic spacings: (a) $x=0$ and (b) $x=999\lambda_a$. The sudden jumps of the probabilities due to retardation are marked by red circles. The first atom is located at $x_1=1\lambda_a$. Note that in the case (b), the atom $``2"$ starts to become excited after a finite time, $t=x/c$, that is due to the retardation.}
\end{figure}

Figure~\ref{population} shows the time evolution of $|b_1(t)|^2$ and $|b_2(t)|^2$ for two different atomic spacings. The frame~(a) illustrates the case when the atoms are very close to each other, with spacing $x\approx 0$. We see that the initially excited atom decays almost exponentially in time, corresponding to the free space regime defined in~\cite{Meystre}. A part of the excitation is transferred directly to the second atom. There is no delay in the excitation of the second atom, as the atomic spacing is negligibly small. A notable feature of the temporal evolution is that at the particular times that correspond to $nL/c$, where $n$ is an integer, a sudden change (jump) in the probabilities occurs. These are just the times when the radiation field emitted into the cavity modes returns to the atoms. It is interesting that the returning radiation does not simultaneously excite both atoms, as one could expect. It rather stimulates a sudden transfer of the population from atom $1$ to atom $2$. 

The sudden jumps continue in time. However, the periodic maxima of the populations are reduced in magnitude as~$t$ increases. This result is consistent with energy-time uncertainty arguments and is readily understood if it is recalled that the excitation wave packet spreads during the evolution, that the excitation becomes less localized as time progresses. An alternative explanation is that there are more and more possible evolution pathways for the excitation to open up as time progresses that are possibly delayed with respect to each other, e.g., due to temporary re-absorptions by the atoms, and then interfere resulting in increased distortions of $|b_1(t)|^2$ and~$|b_2(t)|^2$.

Frame (b) of Fig.~\ref{population} illustrates the time evolution of the probabilities for a large atomic spacing, $x=999\lambda_{a}$. There are now two pathways for the excitation to be transferred between the atoms, $x$ and $L-x$. At very early times, $t \ll L/c$, the initially excited atom $1$ decays with the rate equal to the free space decay rate. The atom $2$ remains in its ground state indicating that initially the excitation is exclusively transferred to the cavity, which essentially appears as open space. The atom $2$ remains in its ground state until the time $t=x/c$, at which the population of the atom $2$ abruptly starts to build up. This is the time required for the excitation emitted by the atom~$1$ to reach the atom $2$ through the shorter pathway $x$. The population of the atom $2$ changes abruptly again at time $t=(L-x)/c$. Note that the abrupt buildup of the excitation of the atom $2$ is not accompanied by an abrupt de-excitation of the atom $1$. There are no sudden changes of the population of the atom $1$ until the time $t=L/c$. This is the time the excitation returns for the first time to the atom~$1$. In fact, neither of the sudden changes of the population of one of the atoms are accompanied by sudden changes of the other. This feature is linked to the fact that the atoms are at different positions and we have taken $x < L/2$. 

Of particular interest is the situation when the two atoms are separated by a distance equal to half of the cavity length. While the sudden kinks in the time evolution of the probability $|b_1(t)|^2$ are still observed at integer multiples of time $L/c$, the number of kinks in evolution of~$|b_2(t)|^2$ reduce to one half as the two paths available for radiation to travel from atom $1$ to atom $2$ are now of equal length. Therefore, in the time behavior of~$|b_2(t)|^2$, kinks are witnessed only at odd integer multiples of $L/(2c)$. Another interesting observation is that there are no sudden jumps of the populations at times $2nx/c$ and $2n(L-x)/c$, where $n$ is an integer, indicating  that the excitation wave packets do not appear to reverse their propagation directions during the interaction with the other atom.

A physical understanding of these behaviors can be obtained if we consider the atomic dynamics in terms of the collective Dicke states of the two-atom system
\begin{eqnarray}
\ket g &=& |g_{1}\rangle\otimes |g_{2}\rangle ,\nonumber\\
\ket s &=& \frac{1}{\sqrt{2}}\left(|e_{1}\rangle\otimes |g_{2}\rangle +|g_{1}\rangle\otimes |e_{2}\rangle \right) ,\nonumber\\
\ket a &=& \frac{1}{\sqrt{2}}\left(|e_{1}\rangle\otimes |g_{2}\rangle - |g_{1}\rangle\otimes |e_{2}\rangle \right) .\label{eq31}
\end{eqnarray}
The advantage of expressing the system in terms of the Dicke state basis is that we can immediately see in which collective state the excitation evolves in time.  

Using Eqs.~(\ref{eq13}) and (\ref{eq31}), we find that the excitation probabilities of the collective symmetric $\ket s$ and the antisymmetric~$\ket a$ states are 
\begin{align}
|b_{s}(t)|^{2} &= \frac{1}{2}\left(|b_{1}(t)|^{2} +|b_{2}(t)|^{2} +2{\rm Re}\left[b_{1}(t)b_{2}^{\ast}(t)\right]\right) ,\nonumber\\
|b_{a}(t)|^{2} &= \frac{1}{2}\left(|b_{1}(t)|^{2} +|b_{2}(t)|^{2} -2{\rm Re}\left[b_{1}(t)b_{2}^{\ast}(t)\right]\right) .
\end{align}

\begin{figure}[h]
\includegraphics[width=0.8\linewidth ]{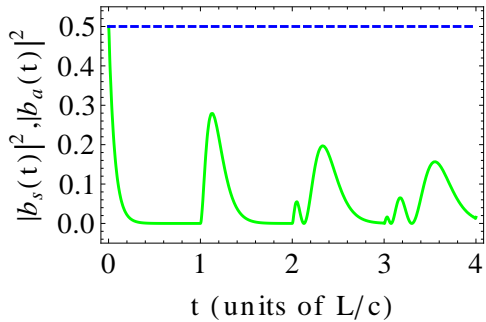}
\caption{(Color online)  Variation of the excitation probabilities $|b_s(t)|^2$ (solid green line) and $|b_a(t)|^2$ (dashed blue line) with time for the same parameters as in Fig.~\ref{population}(a).}
\label{fig4}
\end{figure}

Figure~\ref{fig4} shows how the probabilities $|b_{s}(t)|^{2}$ and $|b_{a}(t)|^{2}$ evolve in time. At $t =0$, the collective states $\ket s$ and $\ket a$ are populated with the same probabilities, $|b_{s}(t)|^{2}=|b_{a}(t)|^{2}=1/2$. The population of the symmetric state decays exponentially in time whereas the population of the antisymmetric state remains constant in time. 
In this figure, the two atoms couple to the cavity modes symmetrically as $x=0$. Therefore, the anti-symmetric excitation state effectively decouples from the cavity, reminiscent of electromagnetically induced transparency or decoherence free sub-spaces. A similar effect can be achieved if the two atoms couple anti-symmetrically to the cavity, in which case the symmetric excitation state remains constant in time.
In contrast, the symmetric state in Fig.~\ref{fig4} becomes re-excited periodically at the time instants given by $nL/c$, where $n$ is an integer. At these times, the emitted radiation field returns to the atoms chronologically. 

We see that the simultaneous sudden changes of both probabilities at the particular discrete times can be explained as an excitation of the collective atomic system from the ground state to the symmetric state. In other words, the jumps represent a collective excitation of the atomic system by the returning radiation field. 

It is interesting to note that shortly before the sudden re-excitation times, the state of the atomic system is 
\begin{align}
|\psi(t=nL/c)\rangle=&\frac{1}{\sqrt{2}}\left(|g\rangle+|a\rangle\right) ,\nonumber\\
=&\frac{1}{\sqrt{2}}|gg\rangle+\frac{1}{2}(|eg\rangle-|ge\rangle),
\end{align}
which shows that the system is in an equal superposition of the ground $\ket g$ and the antisymmetric $\ket a$ states of the two-atom system and explains why in Fig.~\ref{population}(a) $|b_1(nL/c)|^2=|b_2(nL/c)|^2=1/4$.
\begin{figure}[h]
\includegraphics[width=0.8\linewidth ]{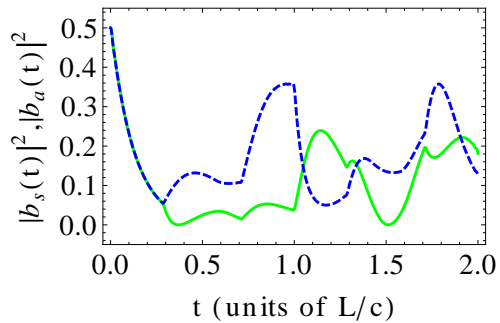}
\caption{(Color online)  Variation of the probabilities $|b_s(t)|^2$ (solid green line) and $|b_a(t)|^2$ (dashed blue line) with time for the same situation as in Fig.~\ref{population}(b).}
\label{fig5}
\end{figure}

Figure~\ref{fig5} shows the time evolution of the populations of the symmetric and antisymmetric states for the same situation as in Fig.~\ref{population}(b). The initial populations decay exponentially with the same rates until $t=x/c$, at which the sudden jump of the populations occurs. A notable difference between the time evolution of $|b_s(t)|^{2}$ and $|b_a(t)|^2$, and that of the individual atoms $|b_1(t)|^{2}$ and $|b_2(t)|^2$, shown in Fig.~\ref{population}(b), is the occurrence of the sudden jumps at the same discrete times. Notice that the most dramatic change in the populations occurs at the time $t=L/c$, i.e. when the excitation returns to the initially excited atom~$1$.

\subsection{Effects of retardation on entanglement - single excitation case}

We now turn to the discussion of the effects of retardation on the entanglement between the atoms. We first focus on short time behavior of the concurrence with two sets of initial conditions in which atoms are prepared in the separable state~(\ref{eq12}) and the maximally entangled state
\begin{equation}
|\psi(0)\rangle = \frac{1}{\sqrt{2}}\left(|e_{1}\rangle\otimes |g_{2}\rangle +|g_{1}\rangle\otimes |e_{2}\rangle \right)\otimes |\{0\}_{\mu}\rangle .\label{eq34}
\end{equation}

The concurrence in the single excitation case can be determined from Eq.~(\ref{Conc.}) in which the probability amplitudes are found solving the set of two coupled equations~(\ref{eq17}).
\begin{figure}[h]
\includegraphics[width=0.8\linewidth ]{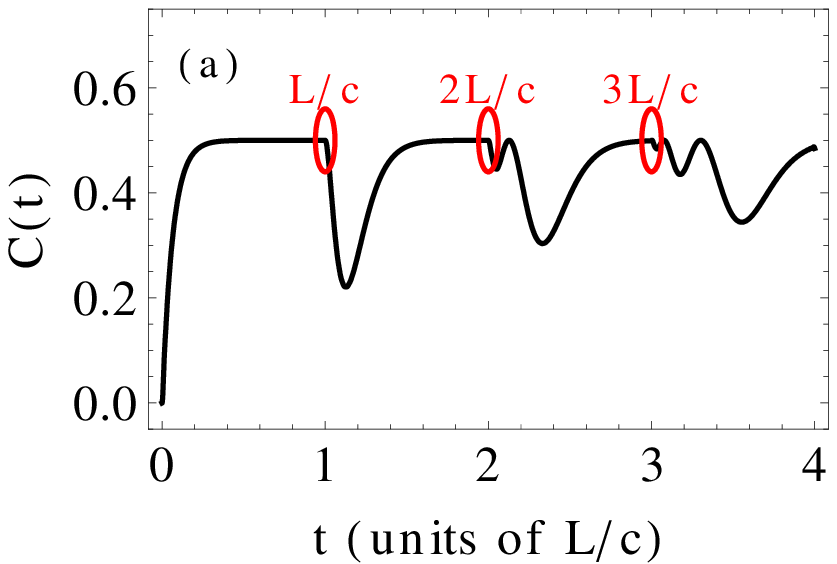}
\includegraphics[width=0.8\linewidth ]{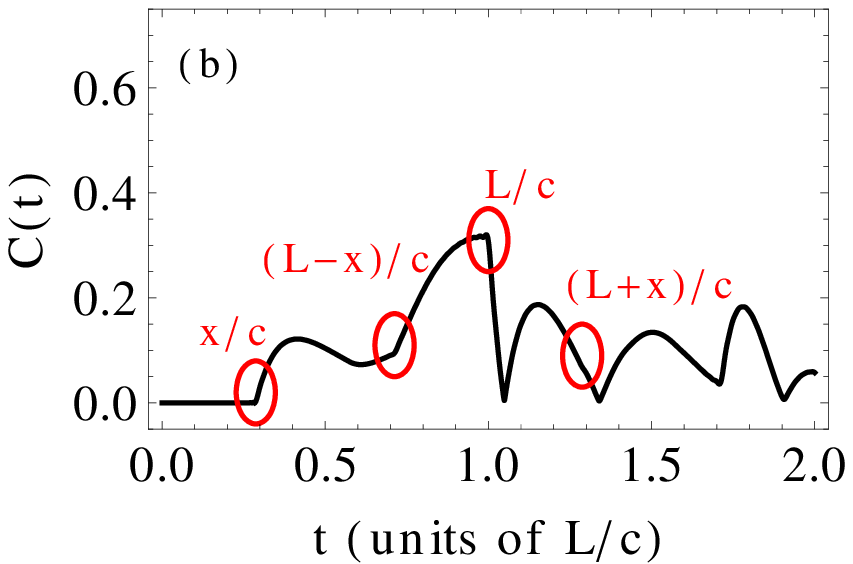}
\caption{(Color online) Transient buildup of entanglement from the initial separable state $\ket{e_{1}g_{2}\{0\}_{\mu}}$ for atomic spacings: (a) $x=0$ and (b) $x=999\lambda_a$. The other parameters are the same as in Fig.~\ref{population}. The red circles mark the positions of the kinks due to retardation.}
\label{fig6}
\end{figure} 

We graph the effect of the retardation on the transient buildup of entanglement between the atoms from the initial separable state (\ref{eq12}) for atomic spacings $x=0$ and $x=999\lambda_{a}$, respectively, in Figs.~\ref{fig6}(a) and~\ref{fig6}(b). We can see how the entanglement between the atoms is affected by the retardation and how it could be related to the population of the collective states.
A comparison of Fig.~\ref{fig6}(a) with Fig.~\ref{fig4} shows that for $x=0$ the manner in which the concurrence evolves in time resembles the evolution of the population of the symmetric state. This is readily understood if one writes the concurrence (\ref{Conc.}) in the basis of the collective Dicke states to find
\begin{equation}
\textbf{C}(t) = {\rm{max}}\left\{0,\sqrt{\left[\rho_{ss}(t)-\rho_{aa}(t)\right]^{2} +\left(2{\rm Im}\left[\rho_{as}(t)\right]\right)^{2}}\right\} .\label{eq35}
\end{equation}
Since $\rho_{aa}(0)=\rho_{aa}(0)=1/2$, and at short times ${\rm Im}[\rho_{as}(t)]\approx 0$, the time evolution of the concurrence depends essentially on the evolution of the population $\rho_{ss}(t)$. It is seen that $\textbf{C}(t)>0$ for all times except $t=0$. It is only at $t=0$ that the atoms are unentangled. The most positive value of $\textbf{C}(t)$ is achieved when $\rho_{ss}(t)=0$, in which case $\textbf{C}(t)=1/2$, so that we may speak of $50\%$ entanglement. The effect of retardation shows up clearly as the sharp decrease of the concurrence from its maximal value of $1/2$. This is due to the transfer of the population from the ground state $\ket g$ to the symmetric state $\ket s$.  

Figure~\ref{fig6}(b) shows the time evolution of the concurrence for a large atomic spacing,  $x=999\lambda_{a}$. The effect of going to a nonzero atomic spacing is clearly to decrease the amount of entanglement and to restrict the time during which it occurs. We see that the initially unentangled atoms remain separable until the time $t=x/c$. The physical reason for the delay in the creation of entanglement is in the retardation effect. No entanglement is created between the atoms until the photon emitted by atom $1$ reaches atom $2$. The atoms remain entangled until the time $t=L/c$ at which the excitation returns to atom $1$. At this time the concurrence suddenly drops to zero. The behavior of the concurrence is entirely consistent with the behavior of the populations of the symmetric and the antisymmetric states, shown in Fig.~\ref{fig5}. 
\begin{figure}[h]
\includegraphics[width=0.8\linewidth ]{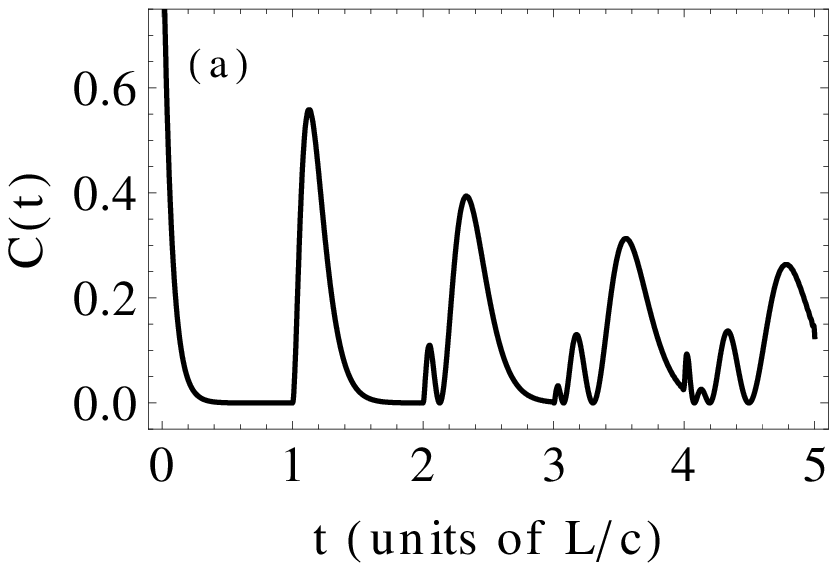}\\
\includegraphics[width=0.8\linewidth ]{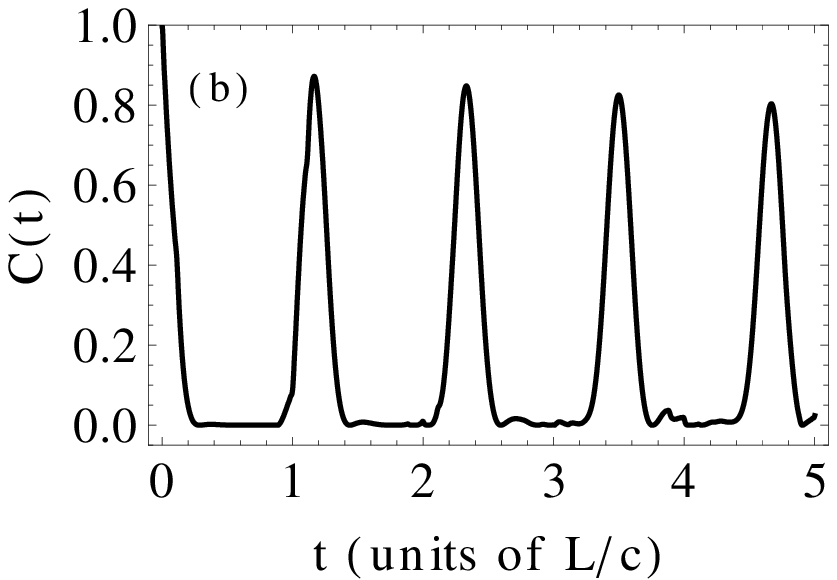}
\caption{Concurrence as a function of time for the same parameters as in Fig.~\ref{population} except that the atoms were initially prepared in the maximally entangled state $\ket s$. The atomic spacing is in (a) $x=0$, and in (b) $x=999\lambda_a$. }
\label{fig7}
\end{figure}

Equation~(\ref{eq35}) predicts that for maximal entanglement between the atoms we would need to put all of the population in one of the collective states, either $\ket s$ or $\ket a$. Following this observation, we plot in  
Fig.~\ref{fig7}(a) the time evolution of the concurrence for the same parameters as in Fig.~\ref{fig6}(a), but with the new initial condition $\rho_{ss}(0)=1$, i.e., the atoms are initially prepared in the maximally entangled state $\ket s$. Since $\rho_{aa}(t)=0$ for $t\geq 0$, the dynamics of the system reduces to that between two states only, the symmetric~$\ket s$ and the ground $\ket g$ states. In this case the concurrence is simply equal to the population of the symmetric state,~$\textbf{C}(t) = \rho_{ss}(t)$. For $t=0$ the atoms are maximally entangled due to our choice of the initial state of the system. Immediately afterwards, the concurrence begins to decrease because of the spontaneous emission to the cavity modes. As soon as the emitted light returns to the atoms, that happens periodically at the times equal to $ nL/c$, where $n$ is an integer, the atoms thereafter become entangled because the system returns to the symmetric state. In the time $t<L/c$, the concurrence approaches zero. This effect, however, is not sudden death of entanglement because $\textbf{C}(t)$ does not become exactly zero. We already found previously that the concurrence for the case of having only one quantum of energy in the system cannot suffer the phenomenon of sudden death because in accordance with Eq.~(\ref{Conc.}), $\textbf{C}(t)$ can either be zero or positive and hence it cannot disappear. In order to have sudden death of entanglement, the second part in the $\rm{max}$ function of Eq.~(\ref{Conc.}) would have to be negative. 

The revival of concurrence at later times can significantly be enhanced by adjusting the atomic spacing. An example is shown in Fig.~\ref{fig7}(b), which is for the same parameters as in frame (a) except for the distance between the atoms. It can be seen that at later times, the time evolution of the concurrence does not split up into multiple peaks as in frame (a). Instead, single peaks with higher amplitudes are obtained. In frame (b), the atomic spacing is chosen such that some retardation revivals coincide with the main concurrence revivals found in frame (a). In particular, the spacing $x$ is adjusted such that the first revival occurs approximately at $t=(L + x)/c$.

\subsubsection{\label{long-single}Long-time dynamics}

In Figs.~\ref{fig6} and \ref{fig7} the concurrence is plotted for short times of the evolution, up to only $t=5L/c$. The results showed that entanglement occurs or is reduced in a periodic fashion, like the pulse periodic excitation, with the magnitude of the subsequent oscillations damped due to the spread of the excitation wave packet. One could expect that the oscillations should collapse after a sufficiently long time and never revive. As we shall see below, this is not the case. Continuing the calculation to much longer times we find that there is an interesting recurrence of the oscillations. 
\begin{figure}[h]
\begin{center}
\includegraphics[width=0.85\linewidth ]{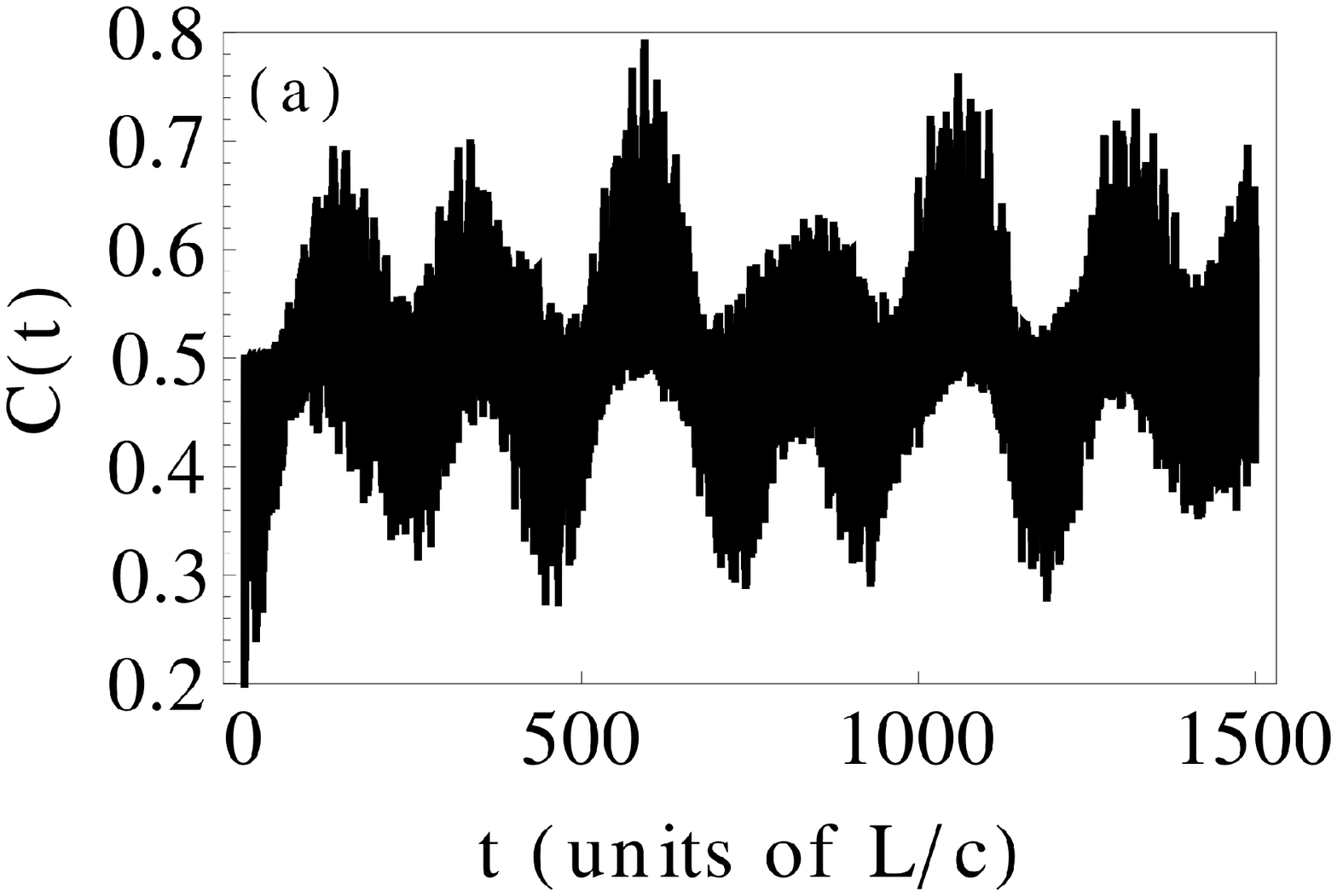}
\includegraphics[width=0.85\linewidth ]{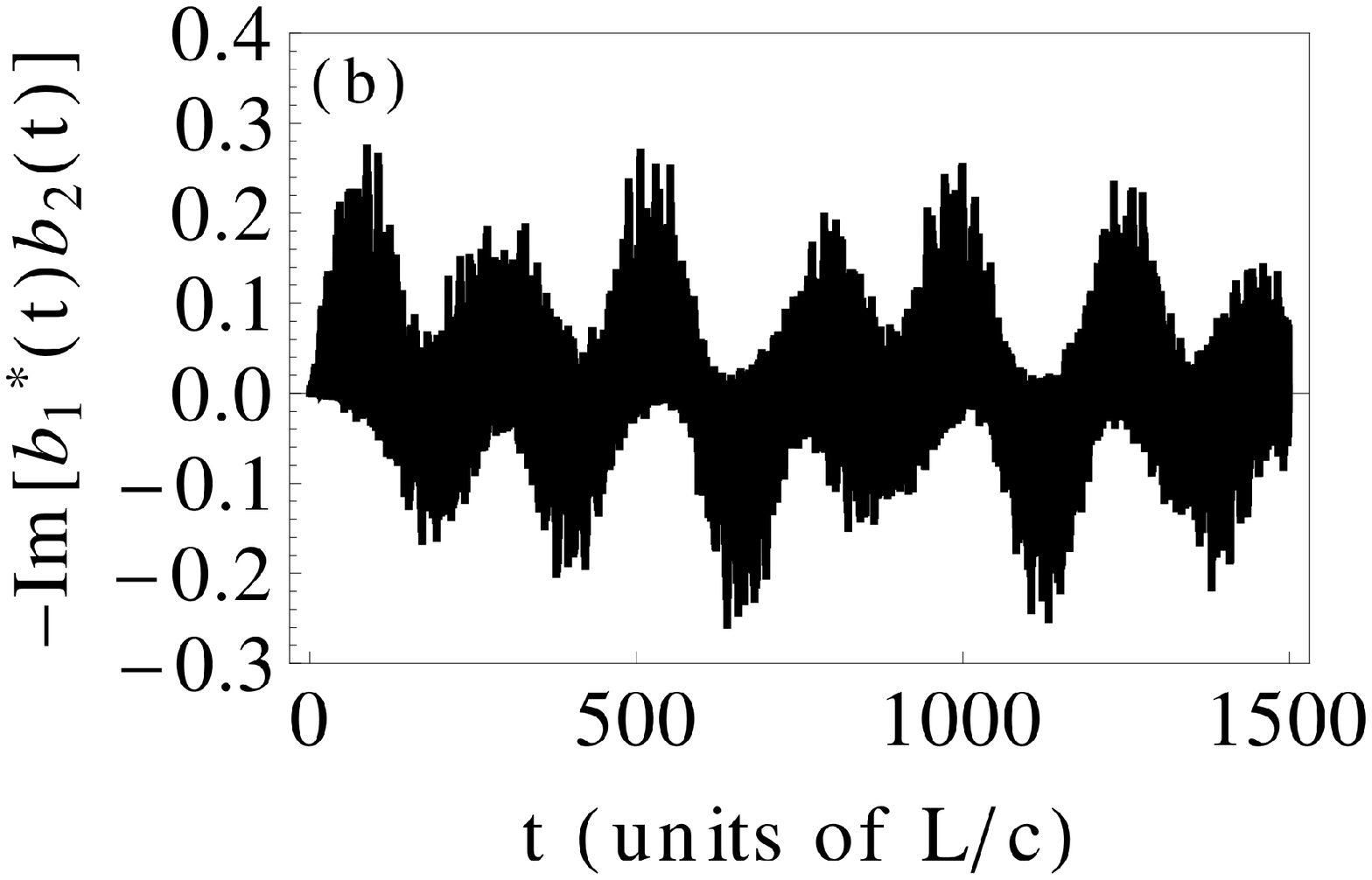}
\caption{\label{longtime}  The long-time behavior of the concurrence for the same situation as in Fig.~\ref{fig6}(a). Frame (a) shows the concurrence $\textbf{C}(t)$, while frame (b) shows the contribution of ${\rm Im}[\rho_{as}(t)]$ to the concurrence, as predicted by Eq.~(\ref{eq35}).}
\end{center}
\end{figure} 

Figure~\ref{longtime}(a) shows the evolution of the concurrence for the same situation as in Fig.~\ref{fig6}(a), but extended to much longer times. It can be seen that the damping of concurrence observed in the initial time evolution does not continue. Rather, on a longer time scale, nearly periodic collapse and revival of the concurrence is observed. Throughout the revivals, the concurrence becomes as large as $\textbf{C}(t) = 0.8$. 

The presence of the pronounced long time oscillations is linked not only to the difference between the populations in the symmetric and anti-symmetric atomic states. Rather, it is also due to an  additional contribution to the concurrence which comes from~${\rm Im}[\rho_{as}(t)]$, see Eq.~(\ref{eq35}). In other words, the coupling of the atoms to the multi mode cavity field leads to a nonzero long-time coherence between the collective states. This is shown in Fig.~\ref{longtime}(b), where we plot ${\rm Im}[\rho_{as}(t)]$ for the same parameters as in frame (a). The coherence is initially zero but beyond $t\sim 10L/c$ starts to build rapidly with the fast oscillations accompanied by a slow modulation. 

\begin{figure}[h]
\begin{center}
\includegraphics[width=0.75\linewidth ]{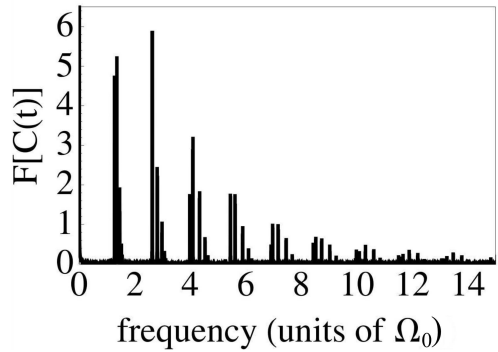}
\caption{\label{fourier} The Fourier transform of $\textbf{C}(t)$ shown in Fig.~\ref{longtime}(a).}
\end{center}
\end{figure} 

The origin of the modulation is in the discrete set of Rabi frequencies $g_{\mu j}$ coupling the atoms to the different modes. The Rabi oscillations are not perfectly periodic due to unequal couplings of the atoms to the discrete modes that causes the imperfection of the modulation. The modulated oscillations bear an interesting relation to the Jaynes-Cummings model with a coherent initial state~\cite{Cummings}. The graininess of the electromagnetic field results in a discrete set of the Rabi frequencies of the coupled atom-field system that are not perfectly periodic but collapse and revive. 

To further analyze the origin of this oscillation, we have calculated the power spectrum of the time signal, which is shown in Fig.~\ref{fourier}. It can be seen that in particular for lower frequencies, the power spectrum decomposes in a set of near-discrete modes. At larger frequencies, the discrete modes decompose into  bands of multiple modes, but the discrete spacing is still visible. This suggest an interpretation of the slow beat-like structure of the long-time dynamics in terms of collapses and revivals, as it is known from the Jaynes-Cummings-model. 

The frequencies appearing in the power spectrum can be traced back to the effective Rabi frequencies occurring in the system of two atoms coupled to many cavity modes. To verify this interpretation, we calculated the time evolution of the atoms analytically in certain limiting cases. The simplest example is the Jaynes-Cummings model~\cite{Cummings}, in which a single atom interacts with a single mode field. Then, the population oscillates at the Rabi frequency $\Omega_{0}$ of the resonant mode which results in a single peak at this frequency in the power spectrum. Similarly, we analyzed the case of two atoms coupling to a single mode, and to two modes. However, in the general case of two atoms coupling to many modes, the analytical calculations become cumbersome and the identification of all the peaks is a complicated task.

\subsubsection{Time-averaged concurrence}

We have seen that the retardation effects show up clearly as the sharp kinks in the concurrence. As seen in Fig.~\ref{longtime}(a), revivals of concurrence appear periodically at long times, with large maximum values of concurrence. But because of the presence of fast oscillations, it is not clear whether the enhancement of the entanglement could be observed in practice. Detectors typically respond over a finite time that could be longer than the oscillation periods of the concurrence. Therefore, we consider the mean concurrence $\langle \textbf{C}(t)\rangle$ averaged over a detection time. As we shall see, the mean concurrence is instructive because it shows how the detected entanglement could be  sensitive to the separation between the atoms. We consider both  long-range and sub-wavelength separations. 
\begin{figure}[h]
\begin{center}
\includegraphics[width=0.8\linewidth ]{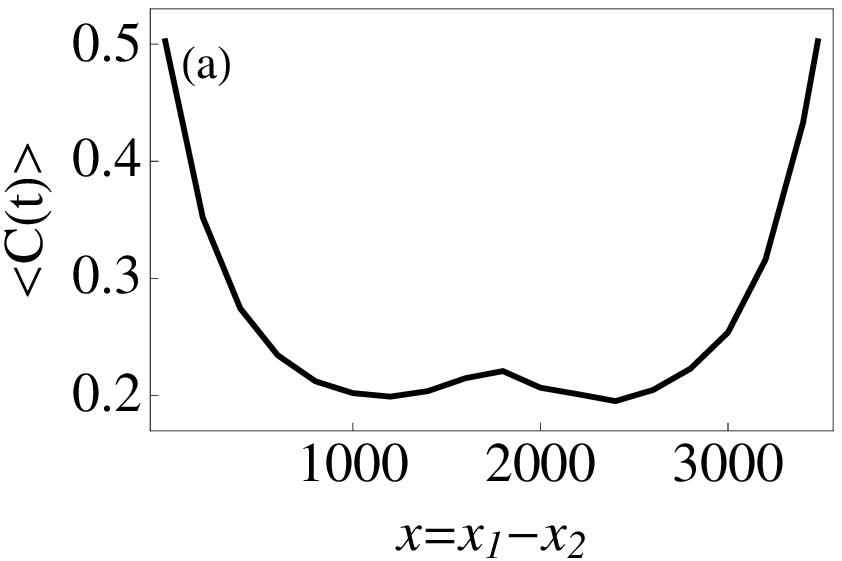}
\includegraphics[width=0.8\linewidth ]{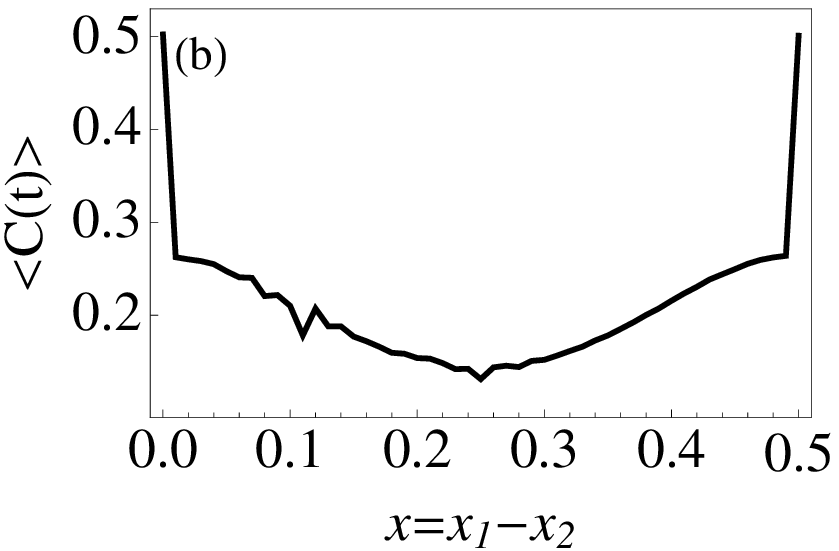}
\caption{\label{mean}  The concurrence averaged over time with respect to the inter atomic separation is shown for the same parameters as in Fig.~\ref{population}. In frame (a) $x$ varies in large steps, while in frame (b) $x$ varies within a half of the wavelength.}
\end{center}
\end{figure}

The mean concurrence $\langle \textbf{C}(t)\rangle$, averaged over a time interval $0\leq t\leq 800L/c$, is shown in Fig.~\ref{mean}.  Frame~(a) illustrates $\langle \textbf{C}(t)\rangle$ at large atomic spacings, $x\gg\lambda_{a}$, with $x$ chosen as integer multiples of $\lambda_a$. Complementarily, frame~(b) shows the variation of $\langle \textbf{C}(t)\rangle$ at sub-wavelength spacings with $x$ varying within a half of the wavelength, $x\leq \lambda_{a}/2$. We observe in both cases that the magnitude of $\langle \textbf{C}(t)\rangle$ is smaller than $1/2$ with the maximum $\langle \textbf{C}(t)\rangle=1/2$ for $x$ at $(0,L)$ for large separations, and at $(0,\lambda_{a}/2)$ for the sub-wavelength separations. Note that the mean concurrence is symmetric with respect to the mid point between the maxima. For the case shown in frame~(a) it corresponds to $x=L/2$, whereas for the case~(b) it corresponds to $x=\lambda_{a}/4$. The behavior of the mean concurrence has a simple explanation. For the separations corresponding to the maxima of the concurrence, the different cavity modes couple to the atoms with the same phases resulting in the same values for the concurrence. For other separations, the atoms experience different phases of the cavity modes relative to the resonant mode, such that the concurrence on average decreases. For the case shown in frame~(a), near one-third of the cavity length, the phase difference among different modes starts decreasing. As a consequence, the curve goes up till the half of the length of the cavity is reached where a symmetry point exists in the sense that the phases of all the even modes match and so do the phases of the odd modes but are completely out of phase from each other. 

We may conclude, that the retardation effects make the concurrence sensitive to the atomic spacing not only at large but also at sub-wavelength spacings.

\subsection{\label{results2}Effects of retardation on entanglement - double excitation case}

We now turn to the discussion of the effects of retardation on the entanglement dynamics when two excitations are present in the system. We show how the well known phenomena resulting from the threshold effects in the concurrence~(\ref{eq26}), such as sudden death, sudden birth and revival of entanglement, can be related to retardation. We will demonstrate that retardation can induce, suppress, or strongly modify these sudden phenomena.
To clearly establish the effect of retardation on the sudden phenomena, we concentrate on properties of the quantity $\mathcal{C}_1(t)$, defined in Eq.~(\ref{eq27}), rather than on~$\mathcal{C}(t)$. Simply speaking, the quantity $\mathcal{C}_1(t)$ can be positive as well as negative which will allow us to distinguish between the sudden phenomena and sudden changes in the evolution due to the retardation that could occur in time periods where the atoms are separable. The concurrence $\mathcal{C}(t)=\mathcal{C}_{1}(t)$ for~$\mathcal{C}_{1}(t)\geq 0$.

We illustrate the role of retardation by examining the time evolution of the system for two sets of initial states for which the sudden phenomena are known to not occur in the absence of retardation. Later, we consider an initial state where even in the absence of retardation, sudden phenomena are present.

\begin{figure}[t]
\includegraphics[width=0.8\linewidth ]{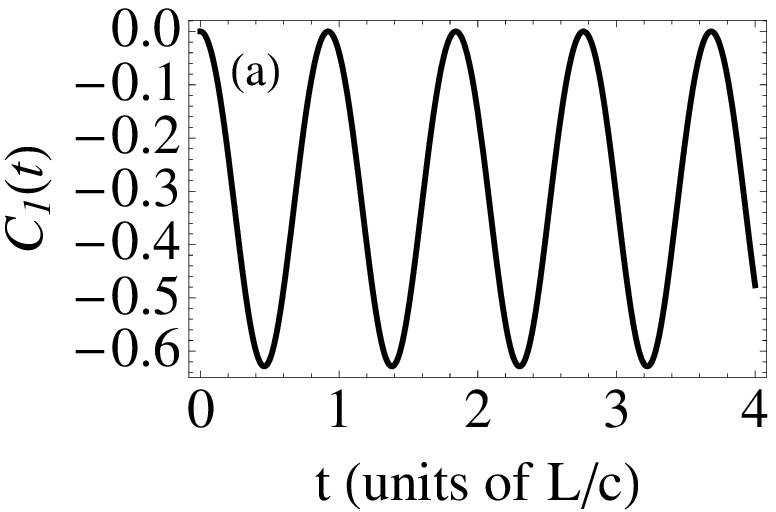}
\includegraphics[width=0.88\linewidth]{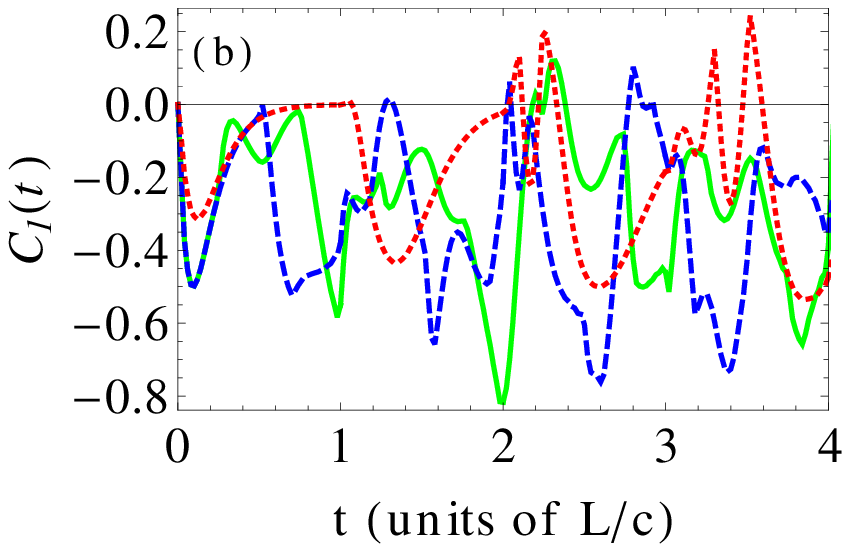}
\caption{(Color online) Time evolution of $\mathcal{C}_1(t)$ for the initial state $\ket{\tilde{\psi}(0)}_{1} = |e_{1}e_{2}\{0\}_{\mu}\rangle$, $L=3.48\times10^3\lambda_a$ and $\omega_a=1.11\times 10^4\Omega_{0}$. In frame (a) $N=1$ and in (b) $N=45$. The dotted red, solid green and dashed blue curves are atomic separations $x=0$, $x=L/4$ and $x=L/2$, respectively. }
\label{sde}
\end{figure}

Consider first an initial state 
\begin{equation}
|\tilde{\psi}(0)\rangle = |e_{1}\rangle\otimes |e_{2}\rangle \otimes |\{0\}_{\mu}\rangle \equiv |e_{1}e_{2}\{0\}_{\mu}\rangle ,\label{eq36}
\end{equation}
in which both atoms are excited and the cavity is empty. Figure~\ref{sde}(a) shows the time evolution of~$\mathcal{C}_1(t)$ when the atoms are coupled to only a single mode $(N=1)$ of the cavity field. In this case no retardation is present. We see that independent of the distance between the atoms, $\mathcal{C}_1(t)$ oscillates sinusoidally in time and is always negative. This indicates that no entanglement is present at any time.

Figure~\ref{sde}(b) shows the corresponding behavior of $\mathcal{C}_1(t)$ for a large number of the cavity modes $(N=45)$ to which the atoms are coupled. In this case the retardation effects occur. It is apparent that the evolution of $\mathcal{C}_1(t)$ is profoundly affected by the presence of retardation. The most interesting aspect of the retardation is the occurrence of the sudden phenomena that lead to an entanglement at some discrete periods of time. The degree of the created entanglement depends on the distance between the atoms.

We have observed similar behavior also for a number of other initial states. Examples are $|\tilde{\psi}(0)\rangle=\ket{\{1\}_{0r}\{1\}_{0l}}\otimes \ket{g_{1}g_{2}}$, that is, none of the atoms is in the excited state and the two photons are in the same (resonant) mode but propagate in opposite directions, or $|\tilde{\psi}(0)\rangle=(1/\sqrt{2})\left(\ket{\{0\}_{0r}\{2\}_{0l}}+\ket{\{2\}_{0r}\{0\}_{0l}}\right)\otimes \ket{g_{1}g_{2}}$, i.e., the atoms are in the ground state, and two photons propagate in the same direction either to the left or to the right in the central cavity mode with equal probability. Again, there is no entanglement if the atoms couple to a single resonant mode of the cavity electromagnetic field. But  entanglement is suddenly born, it suddenly dies and revives in the presence of retardation. 

Consider now an initial state
\begin{equation}
|\tilde{\psi}(0)\rangle = \frac{1}{\sqrt{2}}\left(\ket{\{0\}_{0r}\{1\}_{0l}}+\ket{\{1\}_{0r}\{0\}_{0l}}\right)\otimes \ket{e_{1}g_{2}} ,\label{eq37}
\end{equation}
in which atom $1$ is in excited state, atom $2$ is in ground state and the cavity central counter-propagating modes $\omega_{0r}, \omega_{0l}$ are excited into a coherent superposition of the single-quantum states.

\begin{figure}[h]
\includegraphics[width=0.8\linewidth ]{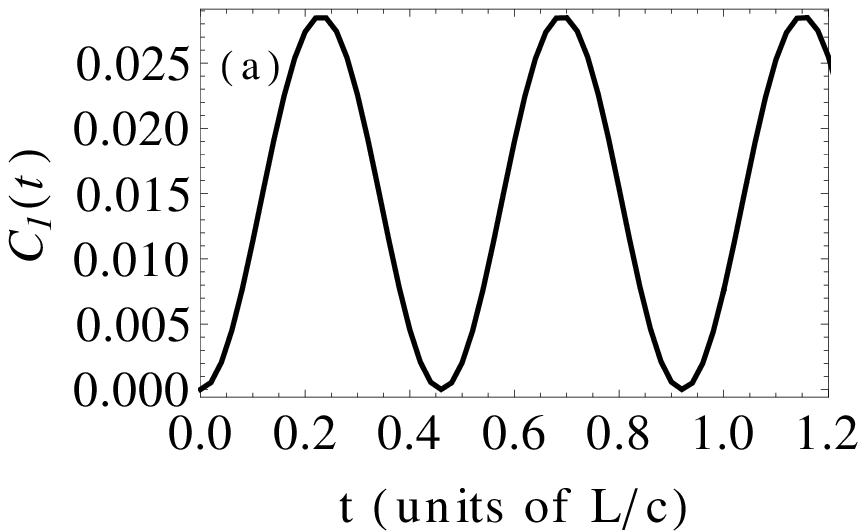}
\includegraphics[width=0.88\linewidth ]{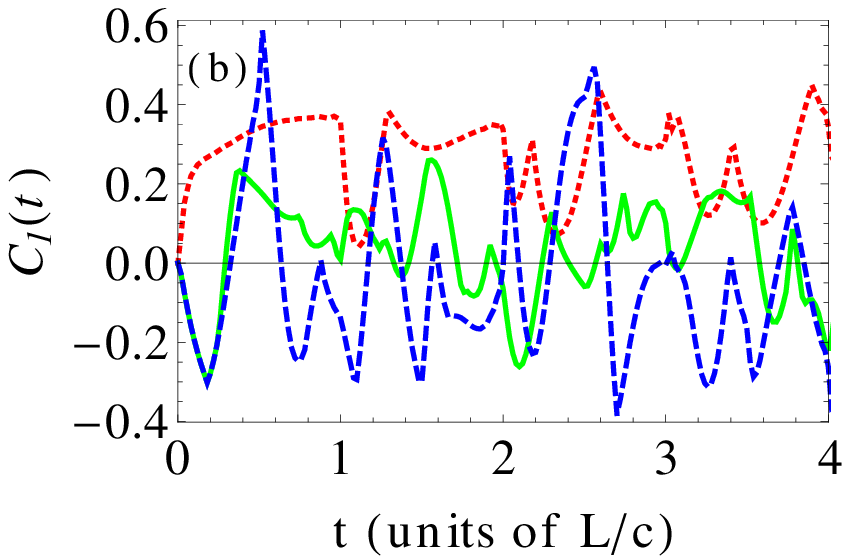}
\caption{(Color online) Time evolution of $\mathcal{C}_1(t)$ for initial condition Eq.~(\ref{eq37}) and the same parameters as in Fig.~\ref{sde}. In frame (a) $N=1$ and in (b) $N=45$. The dotted red, solid green and dashed blue curves correspond to $x=0$, $x=L/4$ and $x=L/2$, respectively.}
\label{SDE7}
\end{figure}

In Fig.~\ref{SDE7}(a) we show the time evolution of $\mathcal{C}_1(t)$ for the initial state (\ref{eq37}) when the atoms are coupled to the central counter-propagating modes. It is seen that $\mathcal{C}_1(t)$ oscillates sinusoidally in time and is non-negative at all times. Once again we notice that no sudden phenomena occur when the atoms are coupled to the central counter-propagating modes. However, contrary to the initial state (\ref{eq36}), the atoms are entangled even when they were initially in a separable state. A naive interpretation for the occurrence of entanglement is that the atoms periodically exchange the excitation through the cavity modes.

The situation becomes different when the atoms couple to a large number of cavity modes. In this case, shown in Fig.~\ref{SDE7}(b), the retardation effects occur and the behavior of~$\mathcal{C}_1(t)$ is seen to be qualitatively different from the previous case. These curves are non-sinusoidal, change sharply in non-periodic way such that $\mathcal{C}_1(t)$ can become negative at some discrete periods of time. Thus, $\mathcal{C}_1(t)$ clearly exhibits the phenomena of sudden death, sudden birth and revival. Again, the degree of concurrence as well as the qualitative dynamics is affected strongly by the inter atomic separation. For example, depending on the distance, atomic entanglement immediately builds up ($d \mathcal{C}_1(0)/dt > 0$), or only at a later time via SBE ($d \mathcal{C}_1(0)/dt < 0$). Interestingly, we find in Fig.~\ref{SDE7}(b) that for negligible separation between the atoms, the concurrence exhibits no death in the presence of retardation, and even persists without intermediate points of zero entanglement in contrast to the non-retarded case. Furthermore, at the separation $x=L/2$, the degree of entanglement is more than one order of magnitude larger than that found in the non-retarded case.

It is interesting to note that the degree of entanglement increases with an increasing separation between the atoms. Again, qualitatively similar behavior is also found for other initial states, such as $(|e_1g_2\{1\}_{0l}\rangle+|g_1e_2\{1\}_{0r}\rangle)/\sqrt{2}$, in which both the atom and the cavity are entangled. 

\begin{figure}
\begin{center}
\includegraphics[width=0.8\linewidth ]{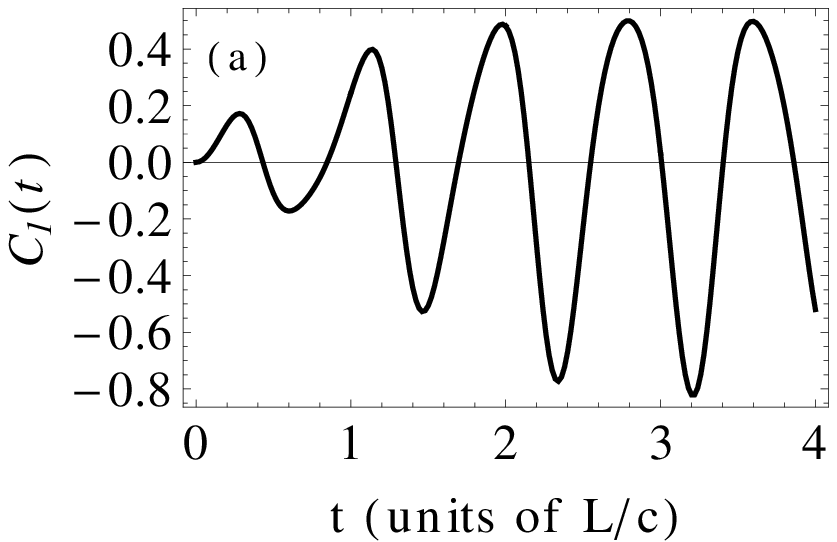}
\includegraphics[width=0.8\linewidth ]{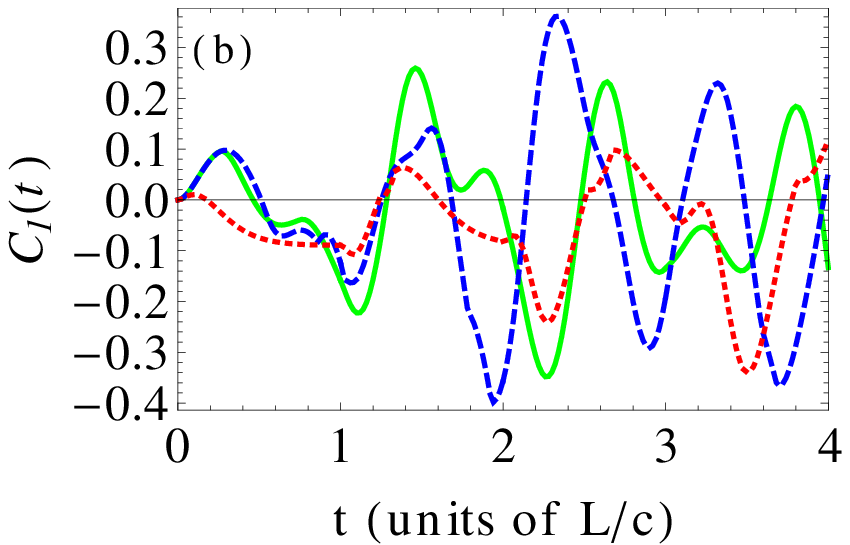}
\caption{\label{SDE2}(Color online) Time evolution of $\mathcal{C}_1(t)$ for the initial state Eq.~(\ref{gg20r}) and the same parameters as in Fig.~\ref{sde}. In frame (a) $N=1$ and in (b) $N=45$. The dotted red, solid green and dashed blue curves correspond to $x=0$, $x=L/4$ and $x=L/2$, respectively.}
\end{center}
\end{figure}

Lastly, we analyze initial states which lead to periodic death and revival of entanglement even without retardation. For this, we consider the separable initial state  
\begin{equation}
 |\tilde{\psi}(0)\rangle = |g_{1}\rangle\otimes |g_{2}\rangle \otimes |\{2\}_{0r}\rangle\equiv |g_{1}g_{2}\{2\}_{0r}\rangle, \label{gg20r}
\end{equation}
in which both atoms are in the ground state, and two photons propagate in the same direction in the central cavity mode. The entanglement dynamics without retardation is shown in Fig.~\ref{SDE2}(a). It can be seen that starting from zero concurrence, entanglement builds up, but then vanishes again. This rebirth and death then repeats periodically. 

The corresponding results with retardation are shown in Fig.~\ref{SDE2}(b). In this case, while the exact temporal dynamics and the magnitude of concurrence is again affected by the inter-particle separation, the qualitative dynamics manifesting itself in the periodic death and birth of entanglement is independent of the retardation effects.

Qualitatively similar results again are also observed for other initial states, such as  $|\tilde{\psi}(0)\rangle=\sqrt{p}|e_1e_2\{0\}_{\mu}\rangle+\sqrt{1-p}|e_1g_2\{1\}_{0l}\rangle$, which we analyzed for $p\in\{1/10,2/10,3/10,4/10\}$. Also the initial state $|\tilde{\psi}(0)\rangle=(|e_1g_2\rangle+|g_1e_2\rangle)|\{1\}_{0l}\rangle/\sqrt{2}$ with maximum entanglement between the atoms behaves qualitatively similar. 

We can gain a qualitative understanding of the behavior of~$\mathcal{C}_1(t)$ in the presence of retardation by making use of Eq.~(\ref{eq31}) for the collective states of the system and expressing~$\mathcal{C}_1(t)$ in terms of the probability amplitudes $b_{\alpha s}(t)$ and~$b_{\alpha a}(t)$ as 
\begin{align}
\mathcal{C}_1(t) &= \sum_{\alpha}\left||b_{\alpha s}(t)|^{2} -|b_{\alpha a}(t)|^{2} -2{\rm Im}\left[b^{\ast}_{\alpha s}(t)b_{\alpha a}(t)\right]\right| \nonumber\\
&-|b(t)|\sqrt{\sum_{\alpha>\beta}|b_{\alpha,\beta}(t)|^2+\sum_{\alpha}|b_{\alpha,\alpha}(t)|^2} , \label{eq38}
\end{align}
where
\begin{align}
b_{\alpha s}(t) &= \frac{1}{\sqrt{2}}\!\left[b_{\alpha1}(t) + b_{\alpha2}(t)\right] ,\nonumber\\
b_{\alpha a}(t) &= \frac{1}{\sqrt{2}}\!\left[b_{\alpha1}(t) - b_{\alpha2}(t)\right] ,\label{eq39}
\end{align}
are the probability amplitudes of the states $\ket s\otimes\ket{\{1\}_{\alpha}}$ and $\ket a\otimes\ket{\{1\}_{\alpha}}$, respectively.

The first line of the right-hand-side of Eq.~(\ref{eq38}) is associated with the one-photon coherence determined by both, unequal populations of the collective states and the coherence between them, whereas the second line is attributable to the two-photon populations of either the atomic system, determined by $b(t)$, or the cavity modes, determined by $b_{\alpha,\beta}(t)$ and $b_{\alpha,\alpha}(t)$. Thus, the mechanism for entanglement with two excitations initially present in the atomic system is similar to that of the single excitation case. Entanglement between the atoms, $\mathcal{C}_1(t)>0$, can be traced back to the single-excitation sub-space, and in particular imbalances between the symmetric and the anti-symmetric singly-excited atom states.  However, in contrast to the single excitation case, this asymmetry must exceed the threshold set by the contribution from the systems with both excitations either in the atoms or in the cavity. 

The entanglement seen in Figs.~\ref{sde}(b),~\ref{SDE7}(b) and ~\ref{SDE2}(b) indicates that the retardation effects lead to a non-zero population difference between the symmetric and antisymmetric states that at some periods of time overcomes the threshold factor in the expression for $\mathcal{C}_1(t)$. What this means is that the time evolution of the atoms is not linked to the total cavity population. In particular, if one excitation is in the cavity, the other excitation can be in different atomic states with varying population imbalance between symmetric and anti-symmetric states.

\subsubsection{Long-time and time-averaged dynamics}

The above investigations have shown that the presence of the retardation effects leads to a non-sinusoidal evolution of the concurrence which results in the phenomena of sudden death and sudden birth of entanglement. One can notice from Figs.~\ref{sde}(b),~\ref{SDE7}(b) and~\ref{SDE2}(b) that there are finite periods of time at which $\mathcal{C}_1(t)$ is negative. These are dead zones of entanglement or equivalently at that times the atoms are separable. The following question then arises: In which states, entangled or separable, do the atoms spend most of the time? To answer this question, we first extend the calculations of $\mathcal{C}_1(t)$ to long times and then average $\mathcal{C}_1(t)$ over a long evolution time. We concentrate on the case illustrated in Fig.~\ref{SDE7}(b).
\begin{figure}[h]
\includegraphics[width=0.8\linewidth ]{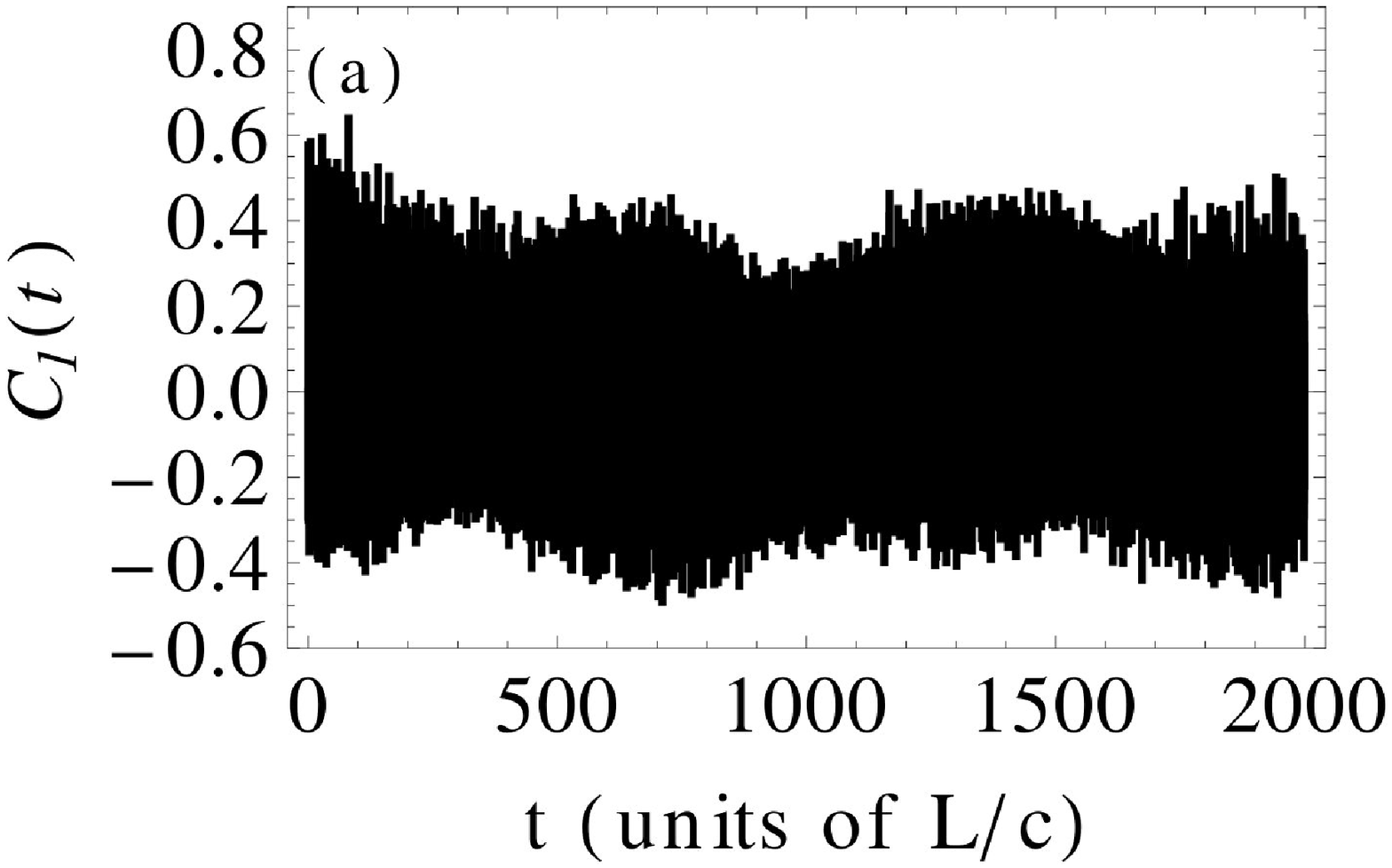}
\includegraphics[width=0.8\linewidth ]{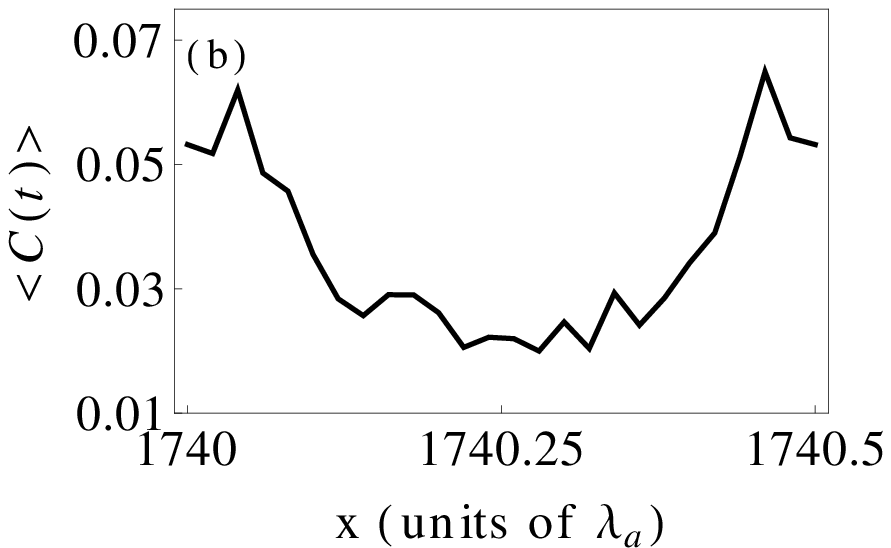}
\caption{ (a) The long time behavior of $\mathcal{C}_1(t)$ is shown for $L=3.48\times10^3\lambda_a, \omega_a=1.11\times 10^4\Omega_{0}, N=45, x=L/2$ and the initial state $|\tilde{\psi}(0)\rangle=(|e_{1}g_{2}\{1\}_{0l}\rangle+|e_{1}g_{2}\{1\}_{0r}\rangle)/\sqrt{2}$.
(b) The concurrence averaged over a long evolution time is shown as a function of the atomic spacing $x$. The distance $x$ varies around half of the cavity length, over an interval of one half of the cavity wavelength.}
\label{result4} 
\end{figure} 

The corresponding long-time dynamics of $\mathcal{C}_1(t)$ is shown in Fig.~\ref{result4}(a). It is seen that the time evolution of $\mathcal{C}_1(t)$ is very spiking with the amplitude of the fast oscillations slowly modulated in time. It can also be noticed that the amplitude of $\mathcal{C}_1(t)$ oscillates around $\mathcal{C}_1(t)=0$ which suggests that over a long period $\mathcal{C}_1(t)$ might average to zero or negative values. Therefore, we calculate the mean value of $\mathcal{C}_{1}(t)$ by averaging over the evolution time $0\leq t \leq 2000L/c$ for different inter atomic distances around a distance of half the cavity length, in an interval of half a wavelength $\lambda_a$. As expected, we find that depending on the inter-atomic separation the mean value of $\mathcal{C}_1(t)$ can be positive or negative. Next, we average the more important concurrence $\mathcal{C}(t)$ itself in the same way. The result is shown in Fig.~\ref{result4}(b). The first atom is located at $x_1=\lambda_a$, and the second atom is located close to $1740\lambda_a$ such that they are separated roughly by half of the cavity length. It is seen that the precise positioning of the atoms within a wavelength at such large distances plays a vital role in the entanglement between the atoms. The effect of going away from the $x=L/2$ position is clearly to decrease the amount of entanglement. But as expected, the concurrence is positive for all distances even in an average sense, due to the non-negative nature of $\mathcal{C}(t)$.

\section{Summary}\label{summary}

We have studied the effects of retardation on the entanglement properties of two atoms located inside a multi-mode ring cavity. Retardation effects become pronounced if the mode spacing of the cavity is small enough such that the atoms can simultaneously couple to many modes of the cavity field. The Schr\"odinger equation for the wave function of the system was solved for different atomic separations and  initial conditions with single and double excitations present in the system. It was shown that the retardation effects are manifest not only in the dynamics of the atomic population but also in the dynamics of entanglement between the atoms.  Characterizing entanglement between the atoms by the concurrence, we have found that the retardation leads to abrupt kinks in the concurrence at intervals corresponding to the flight time of a photon between the atoms or to the time corresponding to a round trip in the cavity. 

Furthermore, we demonstrated that the retardation effects crucially depend on the atom separation both, on the multi- and sub-wavelength distance scale.  We have also distinguished significantly different short-time and long-time retardation effects in the evolution of the concurrence. In particular, at short times the concurrence exhibits periodic sudden changes from separable to highly entangled states. 
At long times, the retardation gives rise to periodic beats in the concurrence that resemble the phenomenon of collapses and revivals in the Jaynes-Cummings model. We finally identified  parameter values and initial conditions at which retardation qualitatively changes the entanglement dynamics. In particular, the atoms can remain either separable or entangled throughout the whole time evolution without retardation, whereas they exhibit the phenomena of sudden birth and sudden death of entanglement when the retardation is included. 

\acknowledgments

QG gratefully acknowledges support by the Higher Education Commission (HEC) of Pakistan administered by Deutscher Akademischer Austauschdienst (DAAD), the International Max Planck Research School (IMPRS) for Quantum Dynamics in Physics, Chemistry and Biology, Heidelberg, Germany, and the Heidelberg Graduate School of Fundamental Physics (HGSFP).

\end{document}